\documentclass[review]{elsarticle}

\usepackage{amsmath}
\usepackage{amsfonts}
\usepackage{amssymb}
\usepackage{amsthm}
\usepackage{bm}
\usepackage{indentfirst}
\usepackage{graphicx}
\usepackage{subfigure}
\usepackage{array}
\usepackage{fullpage}
\biboptions{sort&compress}

\DeclareMathAlphabet{\mathsfsl}{OT1}{cmss}{m}{sl}

\newcommand{\PreserveBackslash}[1]{\let\temp=\\#1\let\\=\temp}
\newcolumntype{C}[1]{>{\PreserveBackslash\centering}p{#1}}
\newcolumntype{R}[1]{>{\PreserveBackslash\raggedleft}p{#1}}
\newcolumntype{L}[1]{>{\PreserveBackslash\raggedright}p{#1}}

\numberwithin{equation}{section}

\theoremstyle{definition}

\usepackage{latexsym,bm}
\usepackage{amssymb}
\usepackage{fancyhdr}
\usepackage{tabularx}
\usepackage{booktabs}
\usepackage{dcolumn}
\usepackage{graphicx}
\usepackage{wrapfig}
\usepackage{multicol}
\usepackage{verbatim}
\usepackage{mathrsfs}
\usepackage{multirow}
\usepackage{tikz}
\usetikzlibrary{calc}
\usepackage{graphicx}
\usepackage{dsfont}
\usepackage{bm}
\usepackage{mathrsfs}
\usepackage{amsmath,mathtools}
\usepackage{amsfonts}
\usepackage{amsthm}
\usepackage{xcolor}
\usepackage{setspace}
\usepackage{float}
\usepackage{url}
\usepackage{multicol}
\usepackage{subfigure}
\usepackage{bm}
\usepackage{multirow}
\usepackage[top=1in,bottom=1in,left=1in,right=1in]{geometry}
\usepackage{diagbox}
\usepackage{algorithm}
\usepackage[noend]{algpseudocode}

\definecolor{officegreen}{rgb}{0.0, 0.5, 0.0}

\makeatletter
\def\ps@pprintTitle{%
  \let\@oddhead\@empty
  \let\@evenhead\@empty
  \let\@oddfoot\@empty
  \let\@evenfoot\@oddfoot
}
\makeatother

\makeatletter
\newcommand*\bdot{\mathpalette\bdot@{.65}}
\newcommand*\bdot@[2]{\mathbin{\vcenter{\hbox{\scalebox{#2}{$\m@th#1\bullet$}}}}}
\makeatother

\makeatletter
\newcommand*\bddot{\mathpalette\bddot@{.65}}
\newcommand*\bddot@[2]{\mathbin{\vcenter{\hbox{\scalebox{#2}
    {$\m@th#1\smash{{}_{\bullet}^{\bullet}}$}}}}}
\makeatother

\newcommand{\real}{\mathbb{R}}

\newcommand{\mcL}{\mathcal{L}}

\newcommand{\mcT}{\mathcal{T}}

\newcommand{\mbR}{\mathbb{R}}

\def\omg{{\Omega}}
\def\omgi{\mathcal{I}{\Omega}}
\def\omgb{\mathcal{B}\Omega}

\def \veps{\varepsilon}

\def \bb{\mathbf{b}}

\def \fb{\mathbf{f}}

\def \ub{\mathbf{u}}
\def \Ub{\mathbf{U}}

\def \xb{\mathbf{x}}

\def \zb{\mathbf{z}}

\def \yb{\mathbf{y}}

\def \Db{\mathbf{D}}
\def \Bb{\mathbf{B}}

\def \Xb{\mathbf{X}}

\newcommand{\verti}[1]{{\left\vert #1
    \right\vert}}  

\begin{document}

\begin{frontmatter}
\title{{{Towards a unified nonlocal, peridynamics framework for the coarse-graining of molecular dynamics data with fractures}}}

\address[yy]{Department of Mathematics, Lehigh University, Bethlehem, PA;}
\address[xx]{Department of Petroleum and Geosystems Engineering, The University of Texas at Austin, Austin, TX;}
\address[ss]{Center for Computing Research, Sandia National Laboratories, Albuquerque,NM;}
\address[md]{Center for Computing Research, Sandia National Laboratories, Livermore, CA;}

\author[yy]{Huaiqian You}
\author[xx]{Xiao Xu}
\author[yy]{Yue Yu\corref{cor1}}\ead{yuy214@lehigh.edu}
\cortext[cor1]{Corresponding author}
\author[ss]{Stewart Silling}
\author[md]{Marta D'Elia}
\author[xx]{John Foster}

\begin{abstract}
{Molecular dynamics (MD) has served as a powerful tool for designing materials with reduced reliance on laboratory testing. However, the use of molecular dynamics directly to treat the deformation and failure of materials at the mesoscale is still largely beyond reach. In this work, we propose a learning framework to extract a peridynamic model as a mesoscale continuum surrogate from MD simulated material fracture data sets. Firstly, we develop a novel coarse-graining method, to automatically handle the material fracture and its corresponding discontinuities in MD displacement data set. Inspired by the Weighted Essentially Non-Oscillatory (WENO) scheme, the key idea lies at an adaptive procedure to automatically choose the locally smoothest stencil, then reconstruct the coarse-grained material displacement field as piecewise smooth solutions containing discontinuities. Then, based on the coarse-grained MD data, a two-phase optimization-based learning approach is proposed to infer the optimal peridynamics model with damage criterion. In the first phase, we identify the optimal nonlocal kernel function from data sets without material damage, to capture the material stiffness properties. Then, in the second phase, the material damage criterion is learnt as a smoothed step function from the data with fractures. As a result, a peridynamics surrogate is obtained. As a continuum model, our peridynamics surrogate model can be employed in further prediction tasks with different grid resolutions from training, and hence allows for substantial reductions in computational cost compared with molecular dynamics. We illustrate the efficacy of the proposed approach with several numerical tests for single layer graphene. Our tests show that the proposed data-driven model is robust and generalizable, in the sense that it is capable in modeling the initialization and growth of fractures under discretization and loading settings that are different from the ones used during training.}
\end{abstract}

\begin{keyword}
Nonlocal Models, Machine Learning, Homogenization, Peridynamics, Material Fracture.
\end{keyword}

\end{frontmatter}


\tableofcontents

\section{Introduction}

Detection and prediction of material damage progression attract lots of interests to the broad scientific and engineering community \cite{zohdi2002toughening,wriggers1998computational,prudencio2013dynamic,su2006guided,AFOSR2014,talreja2015modeling,soric2018multiscale,pijaudier2013damage,mourlas2019accurate,markou2021new}. Physically, the propagation of cracks results from a long-term physical process with its origin in the atomistic scale, which often requires a micro-structural model such as molecular dynamics (MD). However, although MD has made enormous advances in capabilities through better algorithms, better interatomic potentials, and improvements in computational power, its direct employment in treating the deformation and failure of materials at the mesoscale is still largely beyond reach. At the mesoscale and above, a continuum model of mechanics is often required in practice. This fact creates the need for homogenized models that act at larger scales and that, combined with new advanced architectures, allow for fast and accurate predictions of material deformation and fracture \cite{zohdi2017homogenization,bensoussan2011asymptotic,weinan2003multiscale,efendiev2013generalized,junghans2008transport,kubo1966fluctuation,santosa1991dispersive,dobson2010sharp,ortiz1987method,moes1999simplified,hughes2004energy}.

In this paper, we aim to address the question of how to extract coarse-grained measurements and a homogenized surrogate model from MD simulations, that is able to capture material deformation and the nucleation and growth of fractures. Nonlocal models are among the best candidates for this task \cite{du2020multiscale}. In the context of homogenization, nonlocal models are characterized by integral operators (as opposed to differentiable operators) that embed time and length scales in their definition. Therefore, they are able to capture long-range effects that classical PDE models fail to describe \cite{silling2000reformulation}, which makes nonlocal models viable alternatives to partial differential equation (PDE) models when the effects of the small-scale behavior of a system affect its global state \cite{beran70,cher06,karal64,rahali15,smy00,willis85,eringen1972nonlocal,bobaru2016handbook,du2020multiscale}. For monitoring and predicting material fractures, because the nonlocal viewpoint avoids classical notions like deformation gradient, nonlocal models allow a natural description of processes requiring reduced regularity in the relevant solution \cite{bazant2002nonlocal,du2013nonlocal}. As such, the nonlocal continuum mechanics model, in the form of peridynamics \cite{silling_2000,seleson2009peridynamics,parks2008implementing,zimmermann2005continuum,emmrich2007analysis,du2011mathematical,bobaru2016handbook,du2018peridynamic,prakash2016electromechanical,prakash2017computational}, provides a unified modeling of continuum media where continuity and complex material damage modes can be captured autonomously.

In peridynamics and the general nonlocal models, constitutive laws take the form of integrand functions, whose functional form is often justified {\it a posteriori}, which makes rigorous calibration and validation challenging and time consuming. On the other hand, although the nonlocal constitutive laws must be consistent with the classical effective properties, they contain information about the small-scale response of the system and must be chosen to reproduce this response with the greatest fidelity. Therefore, it is desired to extract an optimal integrand function from small-scale data, such that the calibrated nonlocal model reproduces the material responses and can further serve as a homogenized surrogate for future material deformation and fracture prediction tasks \cite{silling2021propagation,silling2022peridynamic}. Recently, with the explosion of machine learning, optimized nonlocal models were designed with the purpose of accurately reproducing observed coarse-grained behavior and predicting unseen behavior with the learnt model. We refer the reader to \cite{You2021,you2020data,you2022data,xu2021machine,xu2022subsurface,zhang2022metanor} for several examples of the use of optimization-based machine learning for the design of homogenized nonlocal operators and the rigorous analysis of its learning theory \cite{lu2022nonparametric,zhang2022metanor}.

Although successful in providing optimal nonlocal surrogates to the homogenization problem, to the authors' best knowledge, none of these approaches addresses the challenge of capturing the main features of dynamic fracture that are seen in small-scale data. Fundamental challenges are still present, mainly due to the two difficulties. Firstly, when mapping the MD measurements onto a coarser grid, coarse graining methods can use the mean atomic velocities weighted by a smoothing function
\cite{murdoch1994continuum}. In a nonlocal setting, a smoothed displacement field can be shown to evolve according to the peridynamic linear momentum balance \cite{you2022data,silling2022peridynamic}. 
However, once the material fracture occurs, such a weighted average approach might overly smooth the displacement field and introduce errors near cracks in the coarse-grained data set. Secondly, in peridynamics the material damage is often described by breaking bonds. Therefore, the integrand functions present jumps near the damage criterion, which results in nonsmooth losses in the optimization problem and hinders the application of a suite of continuous optimization techniques. Herein, we address these two challenges and present a complete workflow demonstrating how to obtain large-scale nonlocal descriptions that capture MD behavior with fractures.

To accomplish this, we develop a novel coarse-graining method which is inspired by the Weighted Essentially Non-Oscillatory (WENO) scheme, and extend the machine learning technique in our previous work \cite{you2022data} to identify a smoothed damage criterion together with optimal nonlocal kernel functions. We summarize below our main contributions.
\begin{itemize}
\item We develop a {\it novel coarse-grained approach from micro-scale fracture measurements}, to automatically choose a locally smoothest stencil and capture the displacement discontinuities. As such, coarse-graining measurements are obtained without overly smoothing the crack pattern.
\item We propose a two-step optimization strategy, and identify the {\it best upscaled surrogate in the form of peridynamics}. Without prior knowledge of the material properties, the resultant peridynamics model describes the material deformation together with the nucleation and growth of fractures.
\item The optimal nonlocal model {\it generalizes well} to fracture patterns that are substantially different from the ones used for training. The optimal model also {\it enables extrapolation to longer time simulations and a multiscale capability to predictions across resolutions.}
\end{itemize}


\paragraph{Outline of the paper}
Section \ref{sec:MD} shows how to obtain an adaptive stencil in the form of a smoothness indicator function, to extract coarse-grained measurements from MD displacements with fractures. In Section \ref{sec:peri} we summarizes the linear peridynamic solid (LPS) model, the treatment of material fracture and the handling of free surfaces, and the discretization technique used in this work. Section \ref{sec:alg} presents our two-step learning approach consisting of a kernel learning step and a damage criterion learning step. Section \ref{sec:resuls} verifies the learning technique for MD displacements and studies the generalizability of the resultant model. On a single layered graphene, we demonstrate efficacy of our workflow by identifying an optimal two-dimensional nonlocal model and employing this model in complex prediction tasks. In particular, we illustrate several properties including generalization with respect to loadings, domain settings, crack shapes, and grid resolutions. Section \ref{sec:conclusion} summarizes our contributions and provides future research ideas.

\section{Coarse-Graining of Molecular Dynamics Displacement with Damage}\label{sec:MD}

\begin{figure}
    \centering
    \includegraphics[width=.48\textwidth]{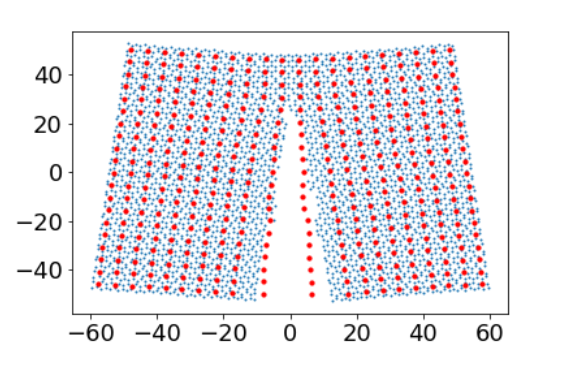}
    \includegraphics[width=.48\textwidth]{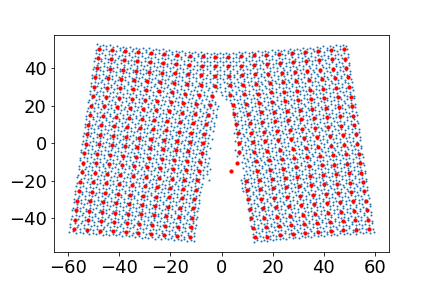}
    \caption{An example of the vanilla coarse-graining method developed in \cite{silling2022peridynamic} and our proposed approach, in handling the MD measurements with a crack. Small blue points represent the MD particles and red dots stand for coarse-grained points. Left: results from the vanilla coarse-graining method with weight function $\omega$. Right: results from our proposed coarse-graining method with an adaptive weight function $\hat{\omega}$.}
    \label{fig:old_coarse}
\end{figure}

\begin{figure}
\centering
\includegraphics[width=1.0\textwidth]{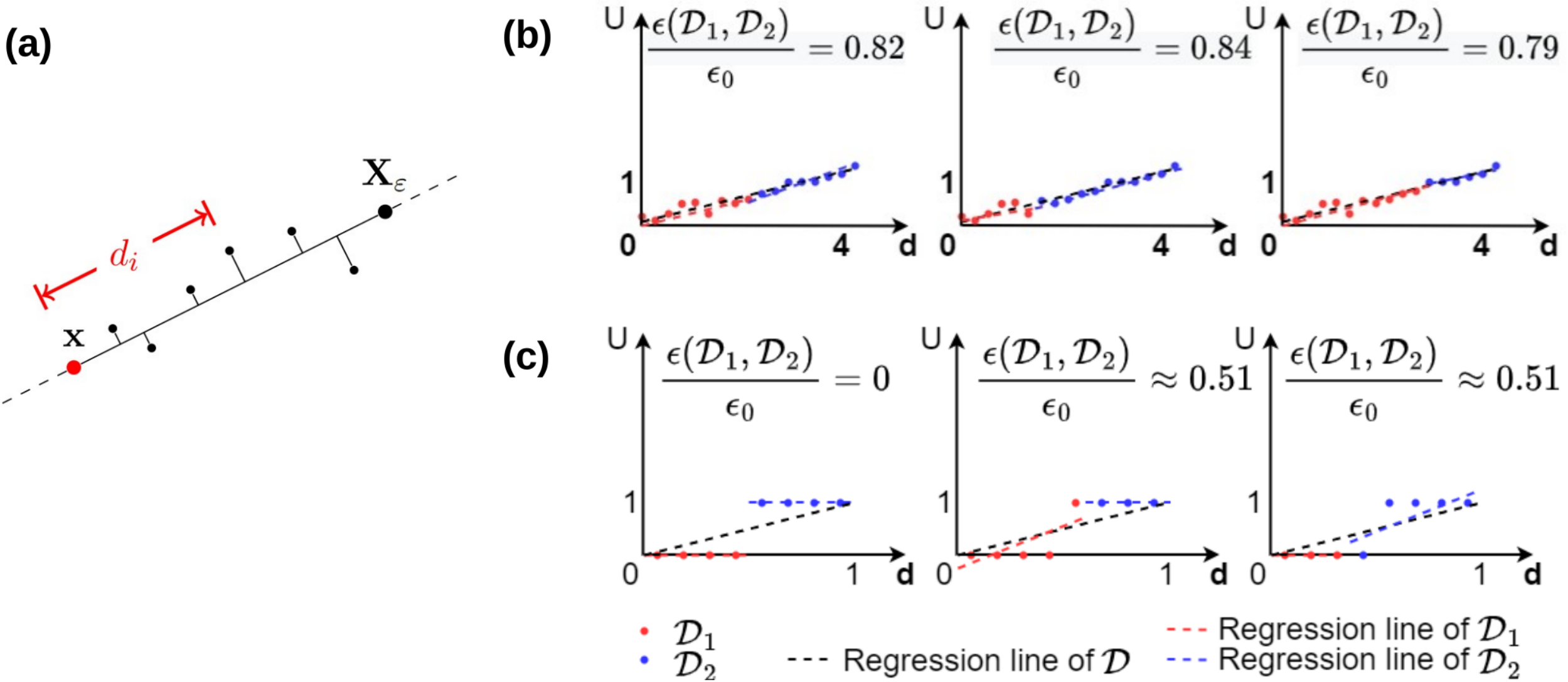}
\caption{Schematic and examples of calculation for the smoothness indicator function $\alpha(\xb, \Xb_{\varepsilon})$. (a): A demonstration of the projection of MD points between $\xb$ and $\Xb_{\varepsilon}$. (b): When the projected displacement field is continuous, the smoothness indicator $\alpha(\xb,\Xb_{\varepsilon})$ stays close to $1$ since $\epsilon(\mathcal{D}_1,\mathcal{D}_2)$ is close to $ \epsilon_0$. (c): when material fracture occurs and the projected displacement field is discontinuous (with a jump at $d = 0.5$), $\dfrac{\epsilon(\mathcal{D}_1,\mathcal{D}_2)}{\epsilon_0}$ reaches the minimum when $\mathcal{D}_1$ and $\mathcal{D}_2$ both consist of a smooth curve, and we have $\alpha(\xb,\Xb_{\varepsilon})\approx 0$.}
\label{fig:smooth_continuous}
\end{figure}

In this section, we introduce the coarse-graining method to map the displacement field data from MD simulations into a larger-scale discretized data cloud. 
For materials without fracture, in \cite{silling2022peridynamic,you2022data} a nonlocal coarse-graining method was proposed. 
In this method, the coarse grained displacement for each particle is defined as a weighted average of the microscale displacements in its neighborhood. As such, a smoothed displacement field is obtained, which preserves a linear momentum balance as a consequence of the momentum balance for the atoms. 

However, this coarse-graining method hides a pitfall: once the material fracture occurs {introducing discontinuities in the displacement field,} the weighted average approach would overly smooth the displacement field and smudge the crack pattern. To resolve this challenge, in this section, we will extend the coarse-graining method to an adaptive approach, so as to automatically handle the material fracture and its corresponding discontinuities in MD displacement data set.




To introduce the coarse-graining method, we consider the MD data set as an assembly of $S$ mutually interacting particles. Then, we define the mass of each mutually interacting particles as 
$M_{\varepsilon}$, $\varepsilon = 1,2,...,S$, the reference positions of
these particles as $\Xb_{\varepsilon}$, and the displacement vectors 
as $\Ub_{\varepsilon}(t)$. Each particle is subjected to a prescribed external force, $\Bb_{\varepsilon}(t)$.

The coarse-grained measurements can be defined by choosing a compactly supported function $\omega(\xb,\cdot)$ for each material point $\xb \in \mathbb{R}^d$, such that 
\begin{equation}\label{eqn:omega}
    \int_{\mathbb{R}^d} \omega(\xb,\Xb_{\varepsilon}) d\xb = 1, \quad 
    \omega(\xb,\Xb_{\varepsilon}) = 0 \quad \text{if} \quad |\xb - \Xb_{\varepsilon}| > R.
\end{equation}
Then, the smoothed material density and body force density are expressed as 
\begin{equation}\label{eqn:density}
    \rho(\xb) = \sum_{\varepsilon=1}^S \omega(\xb,\Xb_{\varepsilon}) M_{\varepsilon},
    \quad 
    \bb(\xb,t) = \sum_{\varepsilon=1}^S \omega(\xb,\Xb_{\varepsilon})
    \Bb_{\varepsilon}.
\end{equation}
Correspondingly, the smoothed displacement field at material point $\xb$ is obtained by 
\begin{equation}
    \ub(\xb,t) = \frac{1}{\rho(\xb)}\sum_{\varepsilon=1}^S \omega(\xb,\Xb_{\varepsilon}) M_{\varepsilon} \Ub_{\varepsilon}(t).
\end{equation}
In \cite{you2022data}, the authors proposed to employ a general cone-shaped weighted function for all material points, by defining
\begin{equation}\label{eqn:omega_smooth}
\omega(\xb_i,\Xb_\varepsilon):=\dfrac{\tau(\xb_i,\Xb_\varepsilon)}{\sum_{j}\tau(\xb_j,\Xb_\varepsilon)},\; \text{ where }\tau(\xb,\Xb)=\max\{0,R-\verti{\Xb-\xb}\}.
\end{equation}
Here, $R$ is a pre-chosen hyperparameter, representing the coarse-graining radius. Such an approach was found to be effective for materials without fracture, where the displacement field is continuous. Then in \cite{you2022data} a data-driven surrogate model was built from this coarse-grained displacement field, which has successfully captured the material properties as well as a constitutive law acting at a larger scale.

However, once the material fracture occurs, the smooth weight function such as the one in \eqref{eqn:omega_smooth} would smudge the crack pattern and hence may compromise the reliability of the resultant surrogate model. As shown in the left plot of Figure \ref{fig:old_coarse}: coarse-grained points appear on the middle of the crack, showing the effect of overly-smoothing the displacement field.

To provide coarse-grained displacement field for both damaged and undamaged material regions, we propose to adjust the smoothing function $\omega (\xb,\Xb_{\varepsilon})$ when the material point $\xb$ is close to the crack. Intuitively, the weight function should choose the locally smoothest stencil and avoid crossing discontinuities in the averaging procedure as much as possible. Similar idea was employed in the Essentially Non-Oscillatory (ENO) and Weighted Essentially Non-Oscillatory (WENO) methods \cite{liu1994weighted,jiang1996efficient}, to develop finite difference schemes for PDE problems with piecewise smooth solutions containing discontinuities. Inspired by the WENO methods, our key idea is to assign an additional smoothness indicator function $\alpha(\xb, \Xb_{\varepsilon})$ to each 
pair of continuum material point $\xb$ and MD particle 
$\Xb_{\varepsilon}$, and modify the weight function $\omega(\xb, \Xb_\varepsilon)$ as:
\begin{equation}
    \hat{\omega}(\xb, \Xb_\varepsilon) = \frac{\omega(\xb, \Xb_\varepsilon)\alpha(\xb,\Xb_{\varepsilon})}{\int_{\mathbb{R}^d} \omega(\xb, \Xb_\varepsilon)\alpha(\xb,\Xb_{\varepsilon}) d\xb}.
\end{equation}
When the crack intersects with the bond between $\xb$ and $\Xb_{\varepsilon}$, the displacement field between $\xb$ and $\Xb_\varepsilon$ contains discontinuity. One should use less information from $\Xb_{\varepsilon}$ to calculate the weighted average on $\xb$, by taking the smoothness indicator function $\alpha(\xb, \Xb_{\varepsilon})\approx 0$. That means, an adaptive procedure is required, to detect if there is a displacement jump between $\xb$ and $\Xb_{\varepsilon}$.

As demonstrated in Fig \ref{fig:smooth_continuous}, we construct the smoothness indicator function $\alpha(\xb, \Xb_{\varepsilon})$ with the following procedure. First, we project all the MD particles within a distance of $R$ from $\xb$ to the line segment that connects $\xb$ and $\Xb_\varepsilon$. For each MD particle $\Xb_i$, we denote the projected point as $\widetilde{\Xb}_i$, and calculate its projected position variable, $d_i$, as the distance from $\xb$ to $\widetilde{\Xb}_i$. The displacement vector $\Ub_i(t)$ on $\Xb_i$ is also projected, and we denote its component along segment $\xb-\Xb_{\varepsilon}$ as $U_i$. Next, we select all the particles with their projections lie between $\xb$ and $\Xb_{\varepsilon}$, to form a set of data pairs $\mathcal{D} = \{(d_i,U_i)\}$. When plotting $U_i$ as a function of $d_i$, the curve will present a jump when there is a crack intersecting the segment between $\xb$ and $\Xb_{\varepsilon}$, and such a jump would naturally divide the set $\mathcal{D}$ as two sets, with each set representing a smooth curve. Therefore, our goal is then to identify the discontinuity of $U(d)$ and define the smoothness indicator function according to it. 
Numerically, we loop over all possible combinations of splitting the data pair set $\mathcal{D}$ into two sets, $\mathcal{D}_1$ and $\mathcal{D}_2$, such that $\mathcal{D}_1\bigcup \mathcal{D}_2=\mathcal{D}$ and $\mathcal{D}_1\bigcap\mathcal{D}_2=\emptyset$. Then, we perform linear regressions on $\mathcal{D}_1$ and $\mathcal{D}_2$:
\begin{equation}
k_\beta,b_\beta=\underset{k,b}{\text{argmin}} \sum_{(d_i,U_i)\in\mathcal{D}_\beta}\verti{kd_i+b-U_i}^2,\quad \beta=1,2,
\end{equation}
to obtain a fitted line for each set. In the meantime, we also perform  a linear regression on the entire displacement data set $\mathcal{D}$, and obtain the fitted parameter set $(k,b)$. Denoting the total squared error $\epsilon(\mathcal{D}_1,\mathcal{D}_2)$ associated with $\mathcal{D}_1$ and $\mathcal{D}_2$ as:
\begin{equation}\label{eqn:linreg}
    \epsilon(\mathcal{D}_1,\mathcal{D}_2) := \sum_{\beta=1}^2\sum_{\{(d_i,U_i)\} \in \mathcal{D}_\beta} (k_\beta d_i + b_\beta - U_i)^2,
\end{equation}
and a squared error for the whole data set $\mathcal{D}$ as
\begin{equation}\label{eqn:linreg_D}
    \epsilon_0 := \sum_{\{(d_i,U_i)\} \in \mathcal{D}} (k d_i + b - U_i)^2,
\end{equation}
we define the smoothness indicator function $\alpha(\xb,\Xb_{\varepsilon})$ as
\begin{align}
    &\alpha(\xb,\Xb_{\varepsilon}) :=  \dfrac{\underset{(\mathcal{D}_1,\mathcal{D}_2)}{\min}\epsilon(\mathcal{D}_1,\mathcal{D}_2)}{\epsilon_0}.
\end{align}
Intuitively, when there is no material fracture and hence $U(d)$ is a smooth curve without discontinuity, we anticipate to have $(k_\beta,b_\beta)\approx(k,b)$ for $\beta=1,2$. As a result, we have $\epsilon(\mathcal{D}_1,\mathcal{D}_2)\approx \epsilon_0$ and $\alpha(\xb,\Xb_{\varepsilon})\approx 1$. In this case, the adjusted weight function $\hat{\omega}(\xb,\Xb_{\varepsilon})$ will stay the same as the original weight function, and hence our smoothness indicator function will not alter the coarse-graining approach for materials without fracture. Figure \ref{fig:smooth_continuous}(b) shows an example with continuous displacement. It can be observed that $\epsilon(\mathcal{D}_1,\mathcal{D}_2)$ stays roughly the same and close to $\epsilon_0$ for different partitions. On the other hand, when fracture occurs, $\epsilon(\mathcal{D}_1,\mathcal{D}_2)$ would achieve its minimum when both $\mathcal{D}_1$ and $\mathcal{D}_2$ consist of a smooth curve. That means, when $\mathcal{D}_1$ and $\mathcal{D}_2$ are separated by the crack. In this case, we will have $\underset{(\mathcal{D}_1,\mathcal{D}_2)}{\min}\epsilon(\mathcal{D}_1,\mathcal{D}_2) < \epsilon_0$ and hence $\alpha(\xb,\Xb_{\varepsilon})< 1$. Therefore, the adjusted weight function $\hat{\omega}(\xb,\Xb_{\varepsilon})$ would automatically reduce the weights of those particle points crossing discontinuities, so as to reduce the overly smoothing near cracks. Figure \ref{fig:smooth_continuous}(c) presents an example where the displacement is piecewise constant with a jump at $d = 0.5$, demonstrating that the smooth indicator would reach its minimum when neither $\mathcal{D}_1$ or $\mathcal{D}_2$ contains the displacement jump.



Once the adjusted weight functions are obtained, the smoothed mass density, body force density, and displacement can be calculated as 
\begin{align}
    \rho(\xb) & = \sum_{\varepsilon=1}^S \hat{\omega}(\xb,\Xb_{\varepsilon}) M_{\varepsilon}, \\ 
    \bb(\xb,t) &= \sum_{\varepsilon=1}^S \hat{\omega}(\xb,\Xb_{\varepsilon})
    \Bb_{\varepsilon}, \\
    \ub(\xb,t) &= \frac{1}{\rho(\xb)}\sum_{\varepsilon=1}^S \hat{\omega}(\xb,\Xb_{\varepsilon}) M_{\varepsilon} \Ub_{\varepsilon}(t).
\end{align}
The result of this modified weight function is demonstrated in the right plot of Figure \ref{fig:old_coarse}. One can see that the coarse-grained points (red dots) are almost aligned with the crack interface, verified the efficacy of our modified coarse-graining method in handling MD data set with cracks.

Similar as the derivation in \cite{you2022data}, we point out that our coarse-grained formulation naturally induces a nonlocal equation of $\ub$. The goal of the present work is therefore to identify an optimal nonlocal model in the form of peridynamics, which faithfully represents given MD displacements under a given set of loading conditions, and is generalizable to further prediction tasks for analysis of material deformation and crack propagation phenomena. 



\section{A Peridynamics Model with Brittle Fracture}\label{sec:peri}

In the previous section, a data set of function trios, $\mcT:=\{(\rho^m,\ub^m,\bb^m)\}_{m=1}^M$, were derived from our coarse-grained formulation, such that each trio contains a coarse-grained density field $\rho^m(\xb)$, a body force density $\bb^m(\xb,t)$, and their corresponding displacement field $\ub^m(\xb,t)$ for material point $\xb\in\omg^m\subset\real^d$ and $t\in[0,T^m]$. Herein, we note that each sample may have different spatial domain $\omg^m$ and observation range $T^m$. We propose to learn a nonlocal momentum balance equation based on these function trios in the form of peridynamic equation of motion \cite{silling_2000}, to provide a continuum model with direct description of fracture within the basic field equations. In peridynamics, each $\xb$ interacts through bond forces with other material points $\yb$ within a neighborhood with radius $\delta$ known as the family of $\xb$, denoted by ${B_\delta(\xb)}$. Here, the horizon $\delta$ determines the extent of the nonlocal interactions. The equation of motion for material point $\xb$ is given as
\begin{equation}
  \rho(\xb)\dfrac{\partial^2\ub(\xb,t)}{\partial t^2}= \int_{{B_\delta(\xb)}}\fb(\yb,\xb,t)\;d\yb+\bb(\xb,t).
    \label{eqn-pdeom}
\end{equation}
A material model in peridynamics supplies values of $\fb(\yb,\xb,t)$ in terms of the deformations of the families of $\xb$ and $\yb$ and any other relevant variables such as temperature \cite{silling_2007}. Peridynamics can model fracture because the equation of motion \eqref{eqn-pdeom} is an integro-differential equation that does not involve the partial derivatives of displacement with respect to position, which leads to a lower requirement of the solution regularity. Moreover, many material models have been developed for peridynamics, and any material model from the local theory can be translated into peridynamic form \cite{silling_2010}. For a more thorough introduction and review about peridynamics, we refer interested readers to \cite{bobaru2016handbook}.

In this work, we aim to addresses the question of how to use MD to obtain a peridynamic material model that is able to treat material deformation and the nucleation and growth of fractures. With a purpose of demonstration and without loss of generality, we consider a 2D simulation problem ($d=2$) in single-layered graphene, and concern small deformations in the linear regime of material response, although the algorithm may be generalized to finite deformations and 3D cases. In \cite{you2022data}, a data-driven two-dimensional linear peridynamic solid (LPS) model\cite{silling_2007} under plane-stress assumption was found to adequately represent the material response from MD simulations on a graphene sheet without fracture. Inspired by such preliminary results, in this work we also employ the LPS model as the base model, and further consider learning of material failure from coarse-grained MD data sets. In this section, we first briefly introduce the LPS model without material fracture in Section \ref{sec:LPS}. Then, we extend the model to describe material fracture and handle the imposition of traction loads as fracture surfaces open up in Section \ref{sec:LPS_damage}. Then, in the next section we will describe our learning algorithm.

\subsection{Linear Peridynamic Solid (LPS) Model}\label{sec:LPS}

In this section, we summarize the mathematical formulation for the LPS model \cite{silling2007peridynamic,emmrich2007well,mengesha14Navier}. The LPS model is a prototypical state-based model which can be seen as a nonlocal extension of the linear elasticity model. It is suitable to describe isotropic elastic materials under infinitesimal deformation. Comparing with the previously developed bond-based peridynamic models \cite{silling_2000,mengesha_du_2014}, the LPS model has advantages in that it is not restricted to a Poisson's ratio of 1/4, which is important for our application since the Poisson ratio of graphene was found to be negative from MD and molecular statistics simulations \cite{qin2017negative,jiang2016intrinsic}.

Consider a body occupying the domain $\Omega\subset\mathbb{R}^d$, and let $\theta$ be the nonlocal dilatation, generalizing the local divergence of the displacement. In this section, we consider the material without damage, with fully prescribed Dirichlet type boundary conditions, and will further extend the discussions to more general boundary conditions and brittle fractures in Section \ref{sec:LPS_damage}. Here we note that in nonlocal problems, unless otherwise stated, the boundary conditions should no longer be prescribed on the sharp interface, $\partial\omg$, but on a collar of thickness of at least $\delta$ surrounding the domain $\omg$, which we denote as:
$$\omgb:=\left\{ \bm x\notin\omg|\text{dist}(\bm x, \partial\omg)  <2\delta\right\}.$$
Given nonlocal boundary conditions prescribed on the nonlocal volumetric boundary domain (or simply nonlocal boundary):
$$\ub(\xb,t) :=  \ub_D(\xb,t)\quad  \xb \in \omgb,\; t\in[0,T],$$
and the initial velocity $\phi(\xb)$ and displacement $\psi(\xb)$ for $\xb\in\omg\bigcup\omgb$ at $t=0$, the peridynamic operator in the LPS model is given by
\begin{equation}\label{eq:nonlocElasticity}
\begin{aligned}
    \mcL_K [\mathbf{u}](\xb,t):=&-\frac{C_1}{m}  \int_{B_\delta (\mathbf{x})} \left(\lambda- \mu\right) K(\left|\mathbf{y}-\mathbf{x}\right|) \left(\mathbf{y}-\mathbf{x} \right)\left(\theta(\mathbf{x},t) + \theta(\mathbf{y},t) \right) d\mathbf{y}\\
  &-  \frac{C_2}{m}\int_{B_\delta (\mathbf{x})} \mu K(\left|\mathbf{y}-\mathbf{x}\right|)\frac{\left(\mathbf{y}-\mathbf{x}\right)\otimes\left(\mathbf{y}-\mathbf{x}\right)}{\left|\mathbf{y}-\mathbf{x}\right|^2}  \left(\mathbf{u}(\mathbf{y},t) - \mathbf{u}(\mathbf{x},t) \right) d\mathbf{y},
  \end{aligned}
\end{equation}
and the nonlocal dilatation is defined via
\begin{equation}\label{eqn:oritheta}
\theta(\mathbf{x},t):=\dfrac{d}{m}\int_{B_\delta (\mathbf{x})} K(\left|\mathbf{y}-\mathbf{x}\right|) (\mathbf{y}-\mathbf{x})\cdot \left(\mathbf{u}(\mathbf{y},t) - \mathbf{u}(\mathbf{x},t) \right)d\mathbf{y},
\end{equation}
where $m:=\int_{B_\delta(\mathbf{0})}K(|\zb|)|\zb|^2d\zb$ is the weighted volume, $\lambda$ is 
Lam\'e's first parameter, and $\mu$ is the shear modulus. To recover parameters for 3D linear elasticity, one should take $C_1=3$, $C_2=30$; whereas for 2D problems, $C_1=2$, $C_2=16$. Here we note that $m$ is determined by the horizon size $\delta$ and the influence function $K$. We use the subscript $K$ in the nonlocal operator $\mcL_K[\ub](\xb)$ to emphasize the operator's dependence on the influence function $K$. Then the time-dependent LPS problem is given by 
\begin{equation}\label{eqn:lps}
    \begin{cases}
    \rho(x)\dfrac{\partial^2{\ub}(\xb,t)}{\partial t^2} + \mcL_K [\mathbf{u}](\xb,t) = \bb(\xb,t), \quad &(\xb,t) \in \Omega\times[0,T]; \\
    \ub(\xb,t) = \ub_{D}(\xb,t), \quad &(\xb,t) \in \omgb\times[0,T]; \\
    \ub(\xb,0) = \psi(\xb), \quad &\xb \in \omg\bigcup\omgb; \\
    \dot{\ub}(\xb,0) = \phi(\xb), \quad &\xb \in \omg\bigcup\omgb. \\
    \end{cases}
\end{equation}
Note that our learning algorithm is compatible with other type of 
boundary conditions in $\omgb$. Here we focus on Dirchlet type of boundary condition in this work for its simplicity.

To discretize the above LPS model, we employ the optimization-based meshfree quadrature rule developed in \cite{trask2019asymptotically,yu2021asymptotically,fan2021asymptotically,you2019asymptotically,you2020asymptotically,foss2021convergence,fan2022obmeshfree,fan2022meshfree}. Suppose that the values of function trios $\rho(\xb)$, $\ub(\xb,t)$, and $\bb(\xb,t)$ are provided on a set\footnote{ Although the machine learning algorithm as well as the quadrature rule is compatible with the general non-uniform grids, in this work we consider the uniform grids with grid size $h$ and uniform time steps with size $\Delta t$, for simplicity.} of coarse-grained material points $\chi:= \{\xb_i\}_{i=1}^{I}\subset \omg\bigcup\omgb$ and time instances $t^n=n\Delta t$, $n=0,\cdots,T/\Delta t$. We write the discretized approximation of $\mcL_{K}$ as 
\begin{equation}\label{eq:discrete}
\begin{aligned}
    \mcL_K^h [\mathbf{u}](\xb_i,t^n):=-\frac{C_1}{m_i}  \sum_{ \mathbf{x}_j \in B_\delta (\mathbf{x_i})\bigcap\chi} \left(\lambda- \mu\right) K_{ij} \left(\mathbf{x}_j-\mathbf{x}_i \right)\left(\theta^h(\mathbf{x}_i,t^n) + \theta^h(\mathbf{x}_j,t^n) \right) W_{j,i}\\
  -  \frac{C_2}{m_i}\sum_{\mathbf{x}_j \in B_\delta (\mathbf{x}_i)\bigcap\chi} \mu K_{ij}\frac{\left(\mathbf{x}_j-\mathbf{x}_i\right)\otimes\left(\mathbf{x}_j-\mathbf{x}_i\right)}{\left|\mathbf{x}_j-\mathbf{x}_i \right|^2}  \left(\mathbf{u}(\xb_j,t^n) - \mathbf{u}(\xb_i,t^n) \right) W_{j,i},
\end{aligned}
\end{equation}
\begin{equation}\label{eqn:disctheta}
\theta^h(\mathbf{x}_i,t^n):=\dfrac{d}{m_i} \sum_{ \mathbf{x}_j \in B_\delta (\mathbf{x}_i)\bigcap\chi} K_{ij} (\mathbf{x}_j-\mathbf{x}_i)\cdot \left(\mathbf{u}(\xb_j,t^n) - \mathbf{u}(\xb_i,t^n) \right)W_{j,i},
\end{equation}
where $K_{ij} := K(\verti{\mathbf{x}_j-\mathbf{x}_i})$ and $m_i:=\underset{\mathbf{x}_j \in B_\delta (\mathbf{x}_i)\bigcap\chi}{\sum}K_{ij}|\mathbf{x}_j-\mathbf{x}_i|^2 W_{j,i}$. The quadrature weights $W_{j,i}$ are associated with a local neighborhood of particles for each discretization point $\xb_i$, generated by local optimizations to make the approximation rule exact for certain classes of functions. For each $\xb_i\in\chi\bigcap\omg$ we solve for $W_{j,i}$ via
\begin{align}\label{eq:quadQP}
  \underset{\left\{\omega_{j,i}\right\}}{\text{argmin}} \sum_{\xb_j \in \chi\bigcap B_\delta(\xb_i)\backslash\{\xb_i\}} \!\!W_{j,i}^2 \quad
  \text{s.t.}, \;
  \sum_{\xb_j\in B_\delta(\xb_i)}\!\!q(\xb_i,\xb_j)W_{j,i} = \int_{B_\delta(\xb_i)} q(\xb_i,\yb)d\yb \quad \forall \,q \in \bm{V}_{\xb_i},
\end{align}
where $\bm{V}_{\xb_i}$ denotes the space of functions which should be 
integrated exactly. Following \cite{yu2021asymptotically}, in this 
work we take $\bm{V}_{\xb_i}:=\left\{q(\yb-\xb_i)= \frac{p(\yb-\xb_i)}{|\yb-\xb_i|^3} \,|\, 
p \in \bm{P}_5(\mathbb{R}^d) \text{ such that } \int_{B_\delta(\mathbf{0})} 
q(\yb) d \yb < \infty \right\}$ and $\bm{P}_5(\mathbb{R}^d)$ denotes 
the space of quintic polynomials. As the horizon size $\delta$ vanishes, this discretization preserves the consistency in the limit to the local solution \cite{trask2019asymptotically,yu2021asymptotically}. Moreover, we point out that the quadrature weights, $W_{j,i}$, only depend on the grid set $\chi$ and it is invariant of the influence $K$. Hence, in our learning algorithm one only need to generate the quadrature weights and solve the local optimization problem \eqref{eq:quadQP} once in the preprocessing step.

For the dynamic peridynamics model, to discretize in time we apply the central difference time stepping scheme. With time step size $\Delta t$, at the $(n+1)-$th time step one can solve for the displacement $\ub_i^{n+1}\approx\ub(\xb_i,t^{n+1})$ following:
\begin{equation}\label{eqn:probdis}
\left\{\begin{array}{ll}
\rho(\xb_i) \ddot{\ub}_{i}^{n} + \mcL_K^h [\mathbf{u}](\xb_i,t^{n}) =  \bb(\xb_i,n\Delta t), & \text{for }\xb_i \text{ in }\Omega\bigcap\chi,\\
\ub_i^{n+1}=\ub_D(\xb_i,(n+1)\Delta t), & \text{for }\xb_i \text{ in }\omgb\bigcap\chi,\\
\end{array}\right.
\end{equation}
where $\mcL_K^h$ is the discretized nonlocal operator as defined in \eqref{eq:discrete}, and the acceleration $\ddot{\ub}_{i}^{n}$ is estimated via the central difference scheme:
\begin{equation}\label{eqn:acc}
    \ddot{\ub}_{i}^{n} := \frac{\ub_i^{n+1} - 2\ub_i^{n} + \ub_i^{n-1}}{\Delta t^2}.
\end{equation}
As the initial conditions, we set $\ub_i^0=\psi(\xb_i)$ and $\dfrac{\ub_i^1-\ub_i^0}{\Delta t}=\phi(\xb_i)$ for $\xb_i \in(\omgb\bigcup\omg)\bigcap\chi$.

\subsection{Peridynamics Formulation for Brittle Fractures}\label{sec:LPS_damage}

One of the main appeals of peridynamics is to handle fracture problems, where free surfaces are associated with the evolution of a fracture surface. In this section, we consider the LPS model with free surfaces, then apply it to the treatment of brittle fractures.

To describe the free surfaces associated with the time evolution of a fracture surface, we now consider general mixed boundary conditions: $\partial\Omega=\partial\Omega_D\bigcup \partial\Omega_N$ %
and $(\partial\Omega_D)^o\bigcap (\partial\Omega_N)^o=\emptyset$. Here $\partial \Omega_D$ and $\partial \Omega_N$ are both curves. $\partial\Omega_N$ is the (possibly time-dependent) sharp crack surface evolving with the material fractures, and a free surface boundary condition is applied on it. To define a Dirichlet-type constraint, we denote
\begin{align*}
\omgb_D:=\{\xb\notin\Omega|\text{dist}(\xb,\partial\Omega_D)<2\delta\},
\end{align*}
and assume that the value of $\ub(\xb,t)=\ub_D(\xb,t)$ is given on $\xb\in\omgb_D$. For notation simplicity, we denote $\omg_D:=\omg\cup\omgb_D$. To apply the free surface boundary condition, we denote
\begin{align*}
\omgi_N&:=\{\xb\in\Omega|\text{dist}(\xb,\partial\Omega_N)<\delta\},\,\omgi:=\{\xb\in\Omega|\text{dist}(\xb,\partial\Omega)<\delta\}.
\end{align*}
Unless stated otherwise, in this paper we further assume sufficient regularity in the boundary region $\omgi$ that there exists a unique orthogonal projection of $\xb$ onto $\partial\Omega$, which is the closest point on $\partial\Omega$ to $\xb$, and we denote this projection as $\overline{\xb}$. Then, one has $\overline{\xb}-\xb=s_x\mathbf{n}({\xb})$ for $\xb\in \omgi_N$, where $0<s_x<\delta$. Here $\mathbf{n}$ denotes the normal direction pointing out of the domain for each $\xb\in\omgi_N$, and let $\mathbf{p}$ denote the tangential direction. In our numerical solver, we treat $\xb$ with the free surface boundary condition if the projection of $\xb$ is in $\partial\omg_N$. Otherwise, we use the Dirichlet-type boundary condition at $\xb$.

\begin{figure}
    \centering
    \includegraphics[width=.6\textwidth]{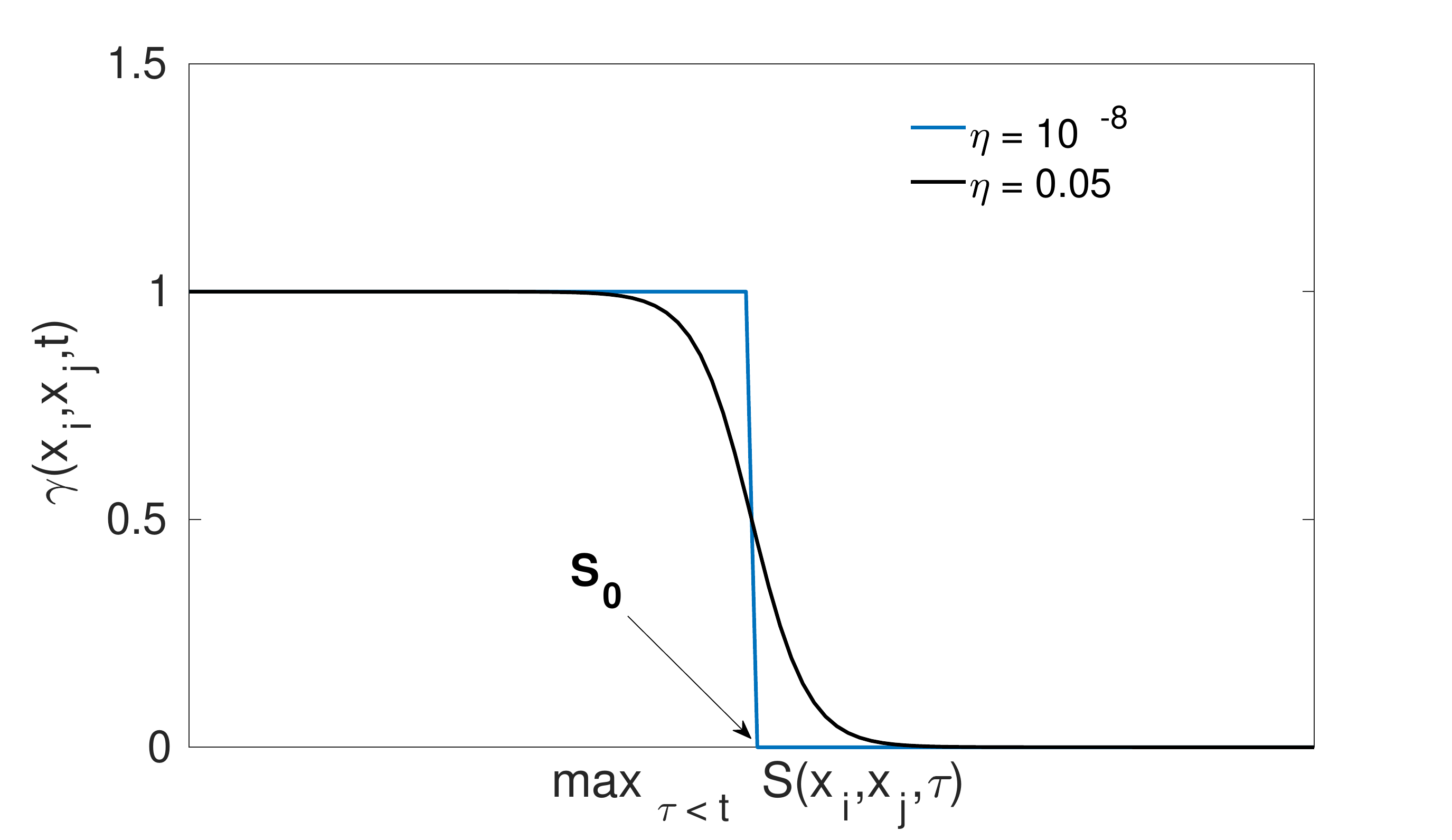}
    \caption{An illustration of the smoothed scalar state function $\gamma$, with the tunable parameter $\eta = \{0.05,10^{-8}\}$.}
    \label{fig:gamma}
\end{figure}

In peridynamics, material damage is incorporated into the constitutive model by allowing the bonds of material points to break irreversibly. To model brittle fracture in the LPS model, we employ a smoothed critical stretch criterion, where weakening occurs when a bond is extended beyond some predetermined critical bond deformed length \cite{zhang2018state,yu2021asymptotically,du2018peridynamic}. In particular, a scalar state function $\gamma(\xb,\yb,t)$ is defined and takes values in the interval $[0,1]$, to describe the bond weakening and breakage through the crack growing:
\begin{equation}\label{eqn:gamma}
\gamma(\xb,\yb,t):=\frac12 \bigg (-\tanh \bigg (\frac{\max_{\tau\in[0,t]}S(\xb,\yb,\tau)-s_0}{\eta} \bigg) + 1 \bigg ),
\end{equation}
where
\begin{equation}
S(\xb,\yb,\tau):=\dfrac{\verti{\xb-\yb+\ub(\xb,\tau)-\ub(\yb,\tau)}}{\verti{\xb-\yb}}-1
\end{equation}
and $s_0$ is the critical stretch criterion depending on the material.  $\gamma(\xb,\yb,t)$ is a history-dependent function, i.e., a bond can never recover once it exceeds the critical stretch criterion. An illustration of $\gamma$ can be visualized in Figure \ref{fig:gamma}, where a hyperparameter $\eta\ll 1$ can be tuned to control the level of smoothness. When $\gamma(\xb,\yb,t)=1$, the bond between material points $\xb$ and $\yb$ are considered ``intact'' and the change of displacement on material point $\yb$ may have an impact on the displacement at $\xb$. When the stretch $S(\xb,\yb,\tau)$ exceeds the critical criterion $s_0$ for some time $\tau<t$, the material gets damaged and we have $\gamma(\xb,\yb,t)<1$. As the stretch further increases, finally $\gamma(\xb,\yb,t)=0$ and we consider the bonds between $\xb$ and $\yb$ as fully ``broken''. Instead of defining $\gamma$ as a step function following \cite{yu2021asymptotically}, in \eqref{eqn:gamma} we allow the weakening of force scalar within small ranges of excessive bond stretch values and set $\gamma$ as a smoothed step function. As shown in \cite{du2018peridynamic}, such a smoothed state function would impose the continuity of the learnt bond force $\fb(\xb,\yb,t)$ in our peridynamic model, and guarantee the well-posedness of the peridynamic model as a dynamic system. On the other hand, a continuous formulation of the damage factor $\gamma$ would result in a continuous optimization problem, and allows generic optimization routines to be used in the training procedure.

With the state function $\gamma$, we treat the time-evolving fracture as free surfaces and employ the following formulation:
 \begin{equation}\label{eqn:probn}
    \begin{cases}
    \rho(x)\dfrac{\partial^2\ub(\xb,t)}{\partial t^2} + \mcL_{KN} [\mathbf{u}](\xb,t) = \bb(\xb,t), \quad &(\xb,t) \in \Omega\times[0,T]; \\
    \ub(\xb,t) = \ub_{D}(\xb,t), \quad &(\xb,t) \in \omgb_D\times[0,T]; \\
    \ub(\xb,0) = \psi(\xb), \quad &\xb \in \omg\bigcup\omgb; \\
    \dot{\ub}(\xb,0) = \phi(\xb), \quad &\xb \in \omg\bigcup\omgb. \\
    \end{cases}
\end{equation}
Here, the modified LPS operator $\mcL_{KN}$ follows the formulation in \cite{yu2021asymptotically,fan2022meshfree}:
 \begin{align}
    \nonumber\mcL_{KN}[\ub]&(\xb,t):=-\frac{C_1}{m}  \int_{B_\delta (\mathbf{x})} \left(\lambda - \mu\right) K(\left|\mathbf{y}-\mathbf{x}\right|)\gamma(\xb,\yb,t)
     \left(\mathbf{y}-\mathbf{x} \right)\left(\theta^{corr}(\mathbf{x},t) + \theta^{corr}(\mathbf{y},t) \right) d\mathbf{y}\\
   \nonumber&-\frac{C_2}{m}\int_{B_\delta (\mathbf{x})} \mu
     K(\left|\mathbf{y}-\mathbf{x}\right|)\gamma(\xb,\yb,t)\frac{\left(\mathbf{y}-\mathbf{x}\right)\otimes\left(\mathbf{y}-\mathbf{x}\right)}{\left|\mathbf{y}-\mathbf{x}\right|^2} 
      \left(\mathbf{u}(\mathbf{y},t) - \mathbf{u}(\mathbf{x},t) \right) d\mathbf{y}\\
    \nonumber&-\frac{2C_1\theta^{corr}(\mathbf{x},t)}{m}  \int_{B_\delta (\mathbf{x})} 
     \left(\lambda - \mu\right) K(\left|\mathbf{y}-\mathbf{x}\right|)(1-\gamma(\xb,\yb,t))
     \left(\mathbf{y}-\mathbf{x} \right) d\mathbf{y}\\
\nonumber&-\frac{C_2\theta^{corr}(\mathbf{x},t)}{2m}\int_{B_\delta (\mathbf{x})}(\lambda+2\mu) K(\left|\mathbf{y}-\mathbf{x}\right|)(1-\gamma(\xb,\yb,t))
     \frac{[\left(\mathbf{y}-\mathbf{x} \right)\cdot \mathbf{n}][\left(\mathbf{y}-\mathbf{x} \right)\cdot \mathbf{p}]^2}{\left|\mathbf{y}-\mathbf{x}\right|^2}
      \mathbf{n} d\mathbf{y} \\
&+\frac{C_2\theta^{corr}(\mathbf{x},t)}{2m} \int_{B_\delta (\mathbf{x})} \lambda K(\left|\mathbf{y}-\mathbf{x}\right|)(1-\gamma(\xb,\yb,t))   \frac{[\left(\mathbf{y}-\mathbf{x} \right)\cdot \mathbf{n}]^3}{\left|\mathbf{y}-\mathbf{x}\right|^2}\mathbf{n} d\mathbf{y}\label{eq:newform1}
 \end{align}
with
\begin{equation}\label{eq:continuousNonlocdilatation3_new}
  \theta^{corr}(\xb,t) := \dfrac{d}{m}\int_{B_\delta (\xb)} K(\left|\yb-\xb\right|)\gamma(\xb,\yb,t) \left(\yb-\xb\right) \cdot \mathbf{M}(\xb)\cdot \left(\ub(\yb,t) - \ub(\xb,t) \right) d\yb,
\end{equation}
\begin{equation}\label{eq:continuousdilCorr_new}
  \mathbf{M}(\xb,t) :=  \left[ \dfrac{d}{m}\int_{B_\delta (\xb)}K(\left|\yb-\xb\right|)\gamma(\xb,\yb,t) \left(\yb-\xb\right) \otimes \left(\yb-\xb\right)  d\yb \right]^{-1}.
\end{equation}
As such, the LPS model provides an approximation for the corresponding linear elastic model with free surfaces in the case of linear displacement fields. We notice that when all bonds in $B_\delta(\xb)$ are intact, i.e., the material point $\xb$ is sufficiently far away from the free surface, we have $\gamma(\xb,\yb,t)=1$ for all $\yb\in B_\delta(\xb)$. Then \eqref{eq:newform1} yields $\mcL_{KN}=\mcL_K$ and the original momentum balance and nonlocal dilatation formulation in the LPS model are obtained. Therefore, \eqref{eq:newform1} provides a unified mathematical framework which automatically captures material deformation and the evolution of cracks as free surfaces.

We now extend the optimization-based quadrature rule and the central difference time-stepping method introduced in Section \ref{sec:LPS}, to the LPS model \eqref{eq:newform1} with fracture. Particularly, at the $(n+1)-$th time step we approximate the state function $\gamma(\xb_i,\xb_j,t^n)$ via
\begin{align}
\gamma^n_{ij} &:= \frac12 \left (-\tanh \left (\frac{\underset{0\leq m\leq n}{\max}S^m_{ij}-s_0}{\eta} \right) + 1 \right ), \text{ where } S^m_{ij}:=\dfrac{|\xb_i-\xb_j+\ub^m_i-\ub_j^m|}
{|\xb_i-\xb_j|}-1.
\end{align}
Then the approximated displacement field $\ub_i^{n+1}\approx\ub(\xb_i,t^{n+1})$ can be solved via the following formulation:
\begin{equation}\label{eqn:probdis_damage}
\left\{\begin{array}{ll}
\rho(\xb_i) \ddot{\ub}_{i}^{n} + \mcL_{KN}^h [\mathbf{u}](\xb_i,t^{n}) =  \bb(\xb_i,n\Delta t), & \text{for }\xb_i \text{ in }\Omega\bigcap\chi,\\
\ub_i^{n+1}=\ub_D(\xb_i,(n+1)\Delta t), & \text{for }\xb_i \text{ in }\omgb_D\bigcap\chi,\\
\end{array}\right.
\end{equation}
where
 \begin{align}
    \nonumber\mcL^h_{KN}[\ub]&(\xb_i,t^{n}):=-\frac{C_1}{m_i}  \underset{\mathbf{x}_j \in B_\delta (\mathbf{x}_i)\bigcap\chi}{\sum} \left(\lambda - \mu\right) K_{ij}\gamma_{ij}^n
     \left(\mathbf{x}_j-\mathbf{x}_i \right)\left((\theta^{corr})_i^n + (\theta^{corr})_j^n \right)W_{j,i}\\
   \nonumber&-\frac{C_2}{m_i}\underset{\mathbf{x}_j \in B_\delta (\mathbf{x}_i)\bigcap\chi}{\sum} \mu
     K_{ij}\gamma_{ij}^n\frac{\left(\mathbf{x}_j-\mathbf{x}_i\right)\otimes\left(\mathbf{x}_j-\mathbf{x}_i\right)}{\left|\mathbf{x}_j-\mathbf{x}_i\right|^2} 
      \left(\mathbf{u}_j^n - \mathbf{u}_i^n \right)W_{j,i}\\
    \nonumber&-\frac{2C_1(\theta^{corr})_i^n}{m_i}  \underset{\mathbf{x}_j \in { B_\delta (\mathbf{x}_i)\bigcap\chi}}{\sum}
     \left(\lambda - \mu\right) K_{ij}(1-\gamma_{ij}^n)
     \left(\mathbf{x}_j-\mathbf{x}_i \right)W_{j,i}\\
\nonumber&-\frac{C_2(\theta^{corr})_i^n}{2m_i}\underset{\mathbf{x}_j \in {B_\delta (\mathbf{x}_i)\bigcap\chi}}{\sum}(\lambda+2\mu) K_{ij}(1-\gamma_{ij}^n)
     \frac{[\left(\mathbf{x}_j-\mathbf{x}_i \right)\cdot \mathbf{n}_i^n][\left(\mathbf{x}_j-\mathbf{x}_i \right)\cdot \mathbf{p}_i^n]^2}{\left|\mathbf{x}_j-\mathbf{x}_i\right|^2}
      \mathbf{n}_i^n W_{j,i}\\
&+\frac{C_2(\theta^{corr})_i^n}{2m_i} \underset{\mathbf{x}_j \in {B_\delta (\mathbf{x}_i)\bigcap\chi}}{\sum} \lambda K_{ij}(1-\gamma_{ij}^n)  \frac{[\left(\mathbf{x}_j-\mathbf{x}_i \right)\cdot \mathbf{n}_i^n]^3}{\left|\mathbf{x}_j-\mathbf{x}_i\right|^2}\mathbf{n}_i^n W_{j,i}\label{eq:newform1_disc}
 \end{align}
with
\begin{equation}\label{eq:continuousNonlocdilatation3_disc}
  (\theta^{corr})_i^n := \dfrac{d}{m_i}\underset{\mathbf{x}_j \in B_\delta (\mathbf{x}_i)\bigcap\chi}{\sum} K_{ij}\gamma_{ij}^n \left(\xb_j-\xb_i\right) \cdot \mathbf{M}_i^n\cdot \left(\ub_j^n - \ub_i^n \right)W_{j,i},
\end{equation}
\begin{equation}\label{eq:continuousdilCorr_disc}
  \mathbf{M}_i^n :=  \left[ \dfrac{d}{m_i}\underset{\mathbf{x}_j \in B_\delta (\mathbf{x}_i)\bigcap\chi}{\sum}K_{ij}\gamma_{ij}^n \left(\xb_j-\xb_i\right) \otimes \left(\xb_j-\xb_i\right)W_{j,i}  \right]^{-1}.
\end{equation}
Here we note that the free surface $\partial\omg_N$ as well as the normal vector $\mathbf{n}({\xb})$ on free surfaces both change as the fracture evolves. To numerically approximate $\mathbf{n}({\xb}_i,t^n)$ at each time step, we updated it via
\begin{equation}\label{eqn:approx_n}
\mathbf{n}^n_{i} = -\dfrac{\underset{\xb_j \in \chi\cap B_\delta(\xb_i)}{\sum}(\xb_j-\xb_i)W_{j,i}\gamma^n_{ij}}{\verti{\underset{\xb_j \in \chi\cap B_\delta(\xb_i)}{\sum}(\xb_j-\xb_i)W_{j,i}\gamma^n_{ij}}}, 
\end{equation}
and the tangential vector $\mathbf{p}^n_{i}$ is calculated as the orthogonal direction to $\mathbf{n}^n_{i}$. The correction tensor should be invertible to ensure that the correction dilitation can be computed. This holds as long as the bonds in the horizon are non-colinear. For fracture case resulting in bond break, leaving an isolated particle, we replace the matrix inverse with the pseudo-inverse.



\section{Learning Algorithm}\label{sec:alg}

\begin{algorithm}
\caption{Workflow for learning the LPS model from MD data sets.}\label{alg:workflow}
\begin{algorithmic}[1]
\State To obtain samples without material fracture, generate relatively small MD displacements on fine grids $\{\Xb_\veps^m\}$ using different external forces and domain configurations, then group the samples into two data sets, $\mathbb{MD}^{\text{Non-Frac}}_{{\text{train}}}$ for training the nonlocal kernel and $\mathbb{MD}^{\text{Non-Frac}}_{{\text{val}}}$ for hyperparameter tuning:
$$\mathbb{MD}^{\text{Non-Frac}}_{{\text{train/val}}}:=\{M^m_{\veps},\Ub^m_{\veps}(t),\Bb^m_{\veps}(t)\},\,m=1,\cdots,M^{\text{Non-Frac}}_{\text{train/val}}.$$
\State
Generate MD displacements samples with material fracture, on fine grids $\{\widetilde{\Xb}_\veps^m\}$ using different external forces and domain configurations, then group the samples into two data sets, $\mathbb{MD}^{\text{Frac}}_{{\text{train}}}$ for training the damage criterion and $\mathbb{MD}^{\text{Frac}}_{{\text{test}}}$ for test:
$$\mathbb{MD}^{\text{Frac}}_{{\text{train/test}}}:=\{\widetilde{M}^m_{\veps},\widetilde{\Ub}^m_{\veps}(t),\widetilde{\Bb}^m_{\veps}(t)\},\,m=1,\cdots,M^{\text{Frac}}_{\text{train/test}}.$$
\State
Coarse grain the data sets $\mathbb{MD}^{\text{Non-Frac}}_{\text{train/val}}$ and $\mathbb{MD}^{\text{Frac}}_{{\text{train/test}}}$, then evaluate the coarse grained data at coarser grids $\chi_m$ to obtain the functio trio sets
$$\mcT^{\text{Non-Frac}}_{\text{train/val}}:=\{\rho^m(\xb_i),\ub^m(\xb_{i},t^n),\bb^m(\xb_{i},t^n)\},\,m=1,\cdots,M^{\text{Non-Frac}}_{\text{train/val}},$$
$$\mcT^{\text{Frac}}_{\text{train/test}}:=\{\widetilde{\rho}^m(\xb_i),\widetilde{\ub}^m(\xb_{i},t^n),\widetilde{\bb}^m(\xb_{i},t^n)\},\,m=1,\cdots,M^{\text{Frac}}_{\text{train/test}}.$$
\State
(\textbf{Kernel learning step}): Solve the optimization problem based on the non-fracture data set $\mcT^{\text{Non-Frac}}_{\text{train}}$:
\begin{flalign*} 
\left\{
\begin{aligned}
(\lambda^*,\mu^*,\Db^*,\alpha^*)=\underset{\lambda,\mu,\Db,\alpha}{\text{argmin }}\; {\rm Res}(\mcT_{\text{train}}^{\text{Non-Frac}}) \\
\text{subject to solvability constraints \eqref{eq:true_problem_full},}
\end{aligned}
\right.
\end{flalign*}
and tune the hyperparameters $\delta^*$ and $P^*$, to minimize the test errors on the validation data set $\mcT^{\text{Non-Frac}}_{\text{val}}$.
\State
(\textbf{Damage criterion learning step}): With fixed parameters $(\lambda^*,\mu^*,\Db^*,\alpha^*,\delta^*,P^*)$, train for the optimal fracture criterion parameter based on the fracture data sets $\mcT^{\text{Frac}}_{\text{train}}$:
\begin{flalign*}
\begin{aligned}
s_0^*=\underset{s_0}{\text{argmin }}\; \widetilde{\rm Res}(\mcT^{\text{Frac}}).
\end{aligned}
\end{flalign*}
\State
To study the generalizability on unseen external forces and fracture scenarios, use the learnt LPS model to predict the material deformation and fracture on $\mcT^{\text{Frac}}_{\text{test}}$.
\end{algorithmic}
\end{algorithm}

Let $\mcT:=\{\rho^m(\xb_{i,m}),\ub^m(\xb_{i,m},t^n_m),\bb^m(\xb_{i,m},t^n_m)\}$, $m=1,\cdots,M$, be coarse-grained function trios available at $\xb_{i,m}\in\chi_m$ and $t^n_m=n\Delta t_m$, $n=1,\cdots,N_m$, our goal is to identify an optimal constitutive relation on the basis of MD data sets. Here, we use $\chi_m$ and $\Delta t_m$ to highlight the fact that in our learning algorithm, each sample can be of different spatial/temporal domain and resolutions. In the following content, we will skip the subscript $m$ and denote the function trios as $\rho^m(\xb_{i})$, $\ub^m(\xb_{i},t^n)$, and $\bb^m(\xb_{i},t^n)$ for simplicity. Let $\mathcal{L}_{KN}$ be the LPS operator defined in \eqref{eq:newform1_disc}, 
we aim to learn an optimal continuum model in the form of LPS models, where the optimal model consists of the influence function $K$, which may be sign-changing, and parameters $\lambda$, $\mu$ and $s_0$, such that the action of $\mathcal{L}_{KN}$ most closely satisfy \eqref{eq:newform1_disc} for all $s$. Formally, the optimal influence function and parameters, $(\lambda^*,\mu^*,s_0^*,K^*)$, are the solution of the following optimization problem:
\begin{equation}\label{abstractOptProblem}
(\lambda^*,\mu^*,s_0^*,K^*) = \underset{\lambda,\mu,s_0,K}{\text{argmin}}{\frac{1}{M}} \sum_{m=1}^M \sum_{n=1}^{N_m-1} \Delta t_m \big\| \rho^m(\xb_{i})(\ddot{\ub}^m)_{i}^n+\mathcal{L}^h_{KN} [\ub^m](\xb_{i},t^n) - \bb^m(\xb_{i},t^n) \big\|^2_{\ell^2(\chi_m)}.
\end{equation}
The influence function $K(\verti{\xb-\yb})$ will now be parameterized. Following \cite{You2020Regression}, In this work, the interacting kernel function $K(\verti{\xb-\yb})$ is taken as a radial function compactly supported on the $\delta$-ball $B_\delta(\mathbf{x})$ with $\alpha$-th order singularity:
\begin{equation}\label{eqn:K}
    K(\verti{\xb-\yb}) = \sum_{k=0}^{P} \frac{D_k}{|\mathbf{x}-\mathbf{y}|^\alpha}B_{k,P}\bigg (\frac{|\mathbf{x}-\mathbf{y}|}{\delta} \bigg).     
\end{equation}
Here the Bernstein polynomials are defined as 
\begin{equation}
    B_{k,P}(r) = \begin{pmatrix}
    P \\
    k \\
    \end{pmatrix}
    r^k(1-r)^{P-k}, \quad \text{ for } 0 \leq r \leq 1.
\end{equation}
Following the arguments in \cite{fan2021asymptotically,you2022data}, in the learning algorithm we require the fractional order $\alpha$ to be bounded by $3$ and allow $D_k\in\mbR$ for all $k$ with sufficient well-posedness conditions embedded for the discretized operator. Here, we note that in the samples with material fracture, some particles might become isolated due to fragmentation, and hence it would be impossible to require solvability constraints. Therefore, we only apply the solvability constraints to the model without fracture. With the analysis in \cite{you2022data}, given a tolerance parameter $\zeta>0$ we apply the following solvability constraints:
\begin{equation}\label{eq:true_problem_full} 
\left\{
\begin{aligned}
\quad &\lambda+\mu>0,\,\mu>0, \, \alpha<3, \,\Lambda_{min}(\Gamma_{(\alpha,\Db,\delta,P)})\geq \zeta,\\
\quad &\Lambda_{min}(\Phi_{(\alpha,\Db,\delta,P)}\Gamma^{\dagger}_{(\alpha,\Db,\delta,P)}\Phi^t_{(\alpha,\Db,\delta,P)})\geq \zeta, \\
\quad &\Lambda_{min}(\Gamma_{(\alpha,\Db,\delta,P)}-2\Phi^t_{(\alpha,\Db,\delta,P)}\Phi_{(\alpha,\Db,\delta,P)})\geq 0.
\end{aligned}
\right.
\end{equation}
Here $\Gamma$ and $\Phi$ are the matrices that correspond to the deviatoric and dilatation contributions of the deformation, and $\Lambda_{min}(A)$ denotes the smallest nonzero eigenvalue of a matrix $A$.

The overall formulation of the constrained optimization problem is as follows. Given a collection of training samples $\{\rho^m(\xb_{i}),\ub^m(\xb_{i},t^n),\bb^m(\xb_{i},t^n)\}$, $m=1,\cdots,M$, we seek to learn the parameters $\lambda$, $\mu$, the Bernstein polynomial coefficients $\Db=[D_0,\cdots,D_P]\in\real^{P+1}$, the order $\alpha$, the horizon $\delta$, the polynomial order $P$, and the damage criterion $s_0$, by minimizing the mean square loss (MSL) of the LPS equation:
\begin{flalign}\label{eq:true_problem} 
\left\{
\begin{aligned}
(\lambda^*,\mu^*,\Db^*,\alpha^*,\delta^*,P^*,s_0^*)=\underset{\lambda,\mu,\Db,\alpha,\delta,P,s_0}{\text{argmin }}\; {\frac{1}{M}} \sum_{m=1}^M \sum_{n=1}^{N_m-1} \Delta t_m &\big\| \rho^m(\xb_{i})(\ddot{\ub}^m)_{i}^n+\mathcal{L}^h_{KN} [\ub^m](\xb_{i},t^n) \\
&- \bb^m(\xb_{i},t^n) \big\|^2_{\ell^2(\chi_m)} \\
\text{subject to solvability constraints \eqref{eq:true_problem_full}.}
\end{aligned}
\right.
\end{flalign}

However, numerically solving the constraint optimization problem \eqref{eq:true_problem_full} could be time-consuming and possibly unstable, due to three factors. First, as shown in Figure \ref{fig:gamma}, when $s_0$ is away from the optimal value, its impact on the loss function would be relatively flattened, causing the vanishing gradient issue in optimizers.  Second, the update of $s_0$ would induce the change of correction operator \eqref{eq:newform1_disc}, which increases the computational cost on each epoch. Lastly, the imposition of solvability constraints \eqref{eq:true_problem_full} would also be expensive, since it involves additional calculations (such as with the projection method) and/or subiterations (such as with the augmented Lagrangian method), together with the evaluation of eigenvalues at each epoch. To make the optimization algorithm more efficient and robust, we propose to separate the solving procedure of the damage criterion, $s_0$, with other parameters, and propose a ``two-stage'' strategy. Key components are summarized in Algorithm \ref{alg:workflow}. In particular, we notice that the correction operator \eqref{eq:newform1_disc} and the damage criterion, $s_0$, are only associated with samples with material fractures, while the influence function $K$ and other material parameters can be inferred from samples without fracture. Therefore, we divide the training data set into two sets:
$$\mcT^{\text{Non-Frac}}:=\{\rho^m(\xb_{i}),\ub^m(\xb_{i},t^n),\bb^m(\xb_{i},t^n)\},\;m=1,\cdots,{M}^{\text{Non-Frac}},$$
which includes all samples without fracture, and
$$\mcT^{\text{Frac}}:=\{\widetilde{\rho}^m(\xb_{i}),\widetilde{\ub}^m(\xb_{i},t^n),\widetilde{\bb}^m(\xb_{i},t^n)\},\;m=1,\cdots,{M}^{\text{Frac}}$$
for training samples with fracture. Then the optimization problem \eqref{eq:true_problem} is also split, into a {\it non-fracture kernel learning} step and a {\it damage criterion learning} step. 

For the kernel learning step, we infer the influence function $K$ and the Lam{\'{e}} moduli $\lambda$ and $\mu$ by solving a constraint optimization problem from $\mcT^{\text{Non-Frac}}$:
\begin{flalign}\label{eq:true_problem_nofrac} 
\left\{
\begin{aligned}
(\lambda^*,\mu^*,\Db^*,\alpha^*,\delta^*,P^*)=\underset{\lambda,\mu,\Db,\alpha,\delta,P}{\text{argmin }}\; {\rm Res}(\mcT^{\text{Non-Frac}}) \\
\text{subject to solvability constraints \eqref{eq:true_problem_full},}
\end{aligned}
\right.
\end{flalign}
where 
\begin{equation}
{\rm Res}(\mcT^{\text{Non-Frac}}):={\frac{1}{{M}^{\text{Non-Frac}}}} \sum_{m=1}^{{M}^{\text{Non-Frac}}} \sum_{n=1}^{N_m-1} \Delta t_m \big\| \rho^m(\xb_{i})(\ddot{\ub}^m)_{i}^n+\mathcal{L}^h_{K} [\ub^m](\xb_{i},t^n) - \bb^m(\xb_{i},t^n) \big\|^2_{\ell^2(\chi_m)}.
\end{equation}
As such, one only has to evaluate the nonlocal operator without fracture following \eqref{eq:discrete}, which is computationally more efficient. In this step, we treat $\delta$ and $P$ as hyperparameters to be separately tuned, to achieve the best learning accuracy without overfitting. For each combination of $\delta$ and $P$, the Adam optimizer in PyTorch is employed, together with the augmented Lagrangian method to impose the inequality constraints. For further details of the optimization algorithm and settings, we refer interested readers to \cite{you2022data}.

In the damage criterion learning step, we fix the learnt parameters $(\lambda^*,\mu^*,\Db^*,\alpha^*,\delta^*,P^*)$ and search for the optimal $s_0$ by considering a unconstraint optimization problem on $\mcT^{\text{Frac}}$:
\begin{flalign}\label{eqn:optim}
\begin{aligned}
s_0^*=\underset{s_0}{\text{argmin }}\; \widetilde{\rm Res}(\mcT^{\text{Frac}}),
\end{aligned}
\end{flalign}
where
\begin{equation}
\widetilde{\rm Res}(\mcT^{\text{Frac}}):={\frac{1}{M^{\text{Frac}}}} \sum_{m=1}^{M^{\text{Frac}}} \sum_{n=1}^{N_m-1} \Delta t_m \big\| \widetilde{\rho}^m(\xb_{i})(\ddot{\widetilde{\ub}}^m)_{i}^n+\mathcal{L}^h_{NK} [\widetilde{\ub}^m](\xb_{i},t^n)- \widetilde{\bb}^m(\xb_{i},t^n) \big\|^2_{\ell^2(\chi_m)}
\end{equation}
In all tests, we set the smoothing parameter $\eta = 0.05$, and employ the bisection method to solve for $s^*_0$.




\section{Application to single-layer graphene}\label{sec:resuls}

To illustrate the capability of our method in obtaining an optimal surrogate material damage model from coarse-grained MD displacements, we consider single layer graphene sheets as the application. Graphene is a single layer of carbon atoms, tightly bound in a hexagonal honeycomb lattice. Up to now, much of what is known about the mechanical and electronic properties of graphene is based on models on the atomistic scale, such as the MD simulations.
However, the use of MD directly to treat the deformation and failure of materials at the mesoscale is still largely beyond reach. Hence, we aim to learn a peridynamic model by upscaling from MD to the continuum scale.

For the present study, an MD model was created using the Tersoff interatomic potential \cite{tersoff88}, a widely used potential in the MD community for graphene \cite{silling2022peridynamic}. Unstressed graphene nominally has an interatomic spacing of 1.46\AA. Without otherwise stated, in this study values of the coarse-grained data trios are evaluated on a square lattice of nodes with spacing $h$=5.0\AA. The only exception is in Section \ref{sec:test_fine}, where we also consider an additional, finer data set generated with spacing 3.17\AA, to assess the generalization properties of the proposed learning approach to different grids. Without the loss of generality, in this work we consider MD simulations on temperature $0K$. In all cases, external loading is applied to the atoms in the MD grid. For the non-fracture data sets, the magnitude of the loading is chosen so that the bond strains are no larger than 2\%, which is less than the strains at which nonlinear effects appear. In all MD experiments, the atoms are initialized with positions on a hexagonal lattice in the $x_1$-$x_2$ plane with an interatomic spacing of $1.46\AA$. The mass of each atom is 2.0E-26kg, or 12amu. For purposes of computing stresses, the thickness of the lattice is set to 3.35\AA, which is the approximate distance between layers in multilayer graphene. On quasi-static data sets, we smooth the MD simulation results in time as described in \cite{you2022data}. For the dynamic data sets, the MD time step size is set as 4.95E-14s, or 49.5fs.


\subsection{Data Generation and Learning results}\label{sec:test_kernel}

\begin{figure}
    \centering
    \includegraphics[width=1.0\textwidth]{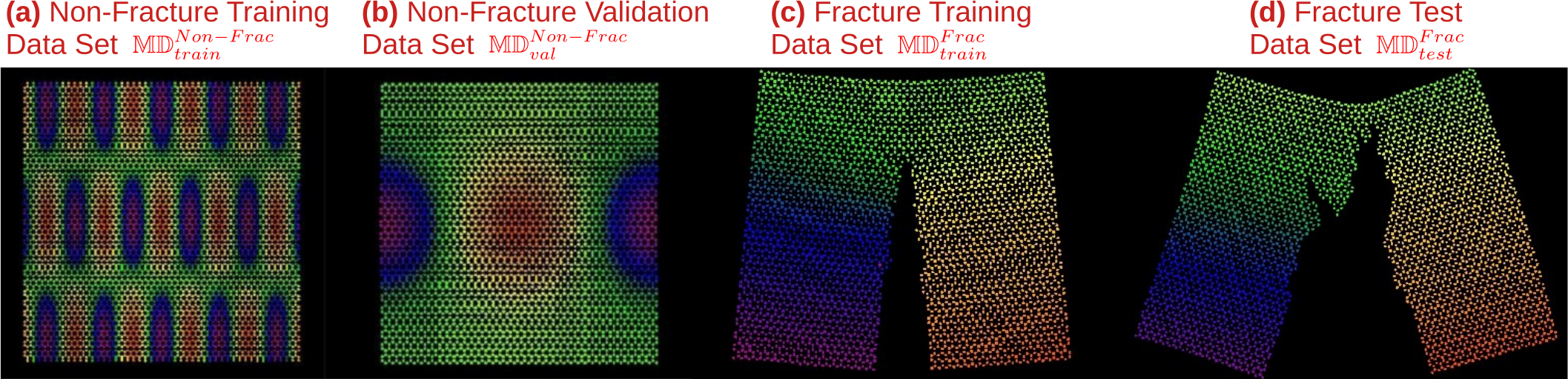}
    \caption{Contours of exemplar $U_1$ displacement in typical MD simulations at zero temperature for the four data sets. From left to right: (a) Non-fracture training data set $\mathbb{MD}^{\text{Non-Frac}}_{{\text{train}}}$, for the kernel learning step; (b) Non-fracture validation data set $\mathbb{MD}^{\text{Non-Frac}}_{{\text{val}}}$ for the kernel learning step; (c) Fracture training data set $\mathbb{MD}^{\text{Frac}}_{{\text{train}}}$, for the damage criterion learning step; (d) Fracture test data set $\mathbb{MD}^{\text{Frac}}_{{\text{test}}}$, to study the efficacy and generalizability of the overall workflow.}
    \label{fig-mddispl}
\end{figure}

In this section, we apply the learning algorithm described in Section \ref{sec:alg}, to extract a coarse-grained model from MD simulations of a graphene sheet at $0K$. 
For the purpose of training, validation and test, we generate the following four groups of MD simulations, with exemplar images showing contours of $U_1$, the component of atomic displacement in the $x_1$ direction, provided in Figure~\ref{fig-mddispl}.

\medskip\noindent
{\it 1) Non-fracture training data set ($\mathbb{MD}^{\text{Non-Frac}}_{{\text{train}}}$, with 70 quasi-static MD simulation samples):} The MD domain is a $100$\AA $\times 100$\AA$\;$ square, and, for {$k_1,k_2\in\{0, \frac{\pi}{50},\frac{2\pi}{50},\ldots,\frac{5\pi}{50}\},$} the prescribed external loadings are given by
\begin{equation}\label{eqn:s_train_300K}
\bb(x_1,x_2) = (C^1_{k_1,k_2}\cos(k_1x_1)\cos(k_2x_2),0), \text{ or } \bb(x_1,x_2) = (0,C^2_{k_1,k_2}\cos(k_1x_1)\cos(k_2x_2)).
\end{equation}
The constant $C^1_{k_1,k_2}$ and $C^2_{k_1,k_2}$ are adjusted so that the bond strains are no larger than 2\%, so the deformation remains in the linear range of material response. A periodic boundary condition is employed for all samples in this data set.

\medskip\noindent
{\it 2) Non-fracture validation data set ($\mathbb{MD}^{\text{Non-Frac}}_{{\text{val}}}$, with 10 quasi-static MD simulation samples).} 
For the same MD grid and coarse-grained nodes as in the non-fracture training data set, the applied loads in the validation data set are as follows:
\begin{equation}\label{eqn:val-load}
\bb(x_1,x_2) = (C_k^1,C_k^2) \sum_{j=-1}^1 (-1)^j 
  \cos\left( \frac{\pi}{2}\min\left\{1,\frac{r_{j,k}}{R_k}\right\} \right)
\end{equation}
where
\begin{equation}\label{eqn:val-rj}
  r_{j,k}=\sqrt{ (x_1-(1-p_k)Lj)^2+(x_2-p_kLj)^2 }
\end{equation}
where $L$=50
and the values of the parameters $C_k^1$, $C_k^2$, $p_k$ and $R_k$ are given in Table~\ref{table-valparam}.
In each case, loads are applied to the atoms within three disks of radius $R_k$ with centers at the
center of the grid and at the left and right boundaries (if $p_k=0$) or the upper and lower boundaries (if $p_k=1$). The loads in all cases are self-equilibrated and periodic.

\begin{table}  [t]
\centering
    \begin{tabular}{|c|c|c|c|c|} \hline
        $k$ & $C_k^1$     & $C_k^2$  & $p_k$ & $R_k$ \\  \hline\hline
        1   & 0.001       & 0        & 0     & 25    \\  \hline
        2   & 0           & 0.001    & 0     & 25    \\  \hline
        3   & 0.001       & 0        & 0     & 15    \\  \hline
        4   & 0           & 0.001    & 0     & 15    \\  \hline
        5   & 0.001       & 0        & 0     & 10    \\  \hline
        6   & 0.001       & 0        & 1     & 25    \\  \hline
        7   & 0           & 0.001    & 1     & 25    \\  \hline
        8   & 0.001       & 0        & 1     & 15    \\  \hline
        9   & 0           & 0.001    & 1     & 15    \\  \hline
        10  & 0.001       & 0        & 1     & 10    \\  \hline
    \end{tabular}
    \caption{Parameters used in the MD loading in the 10 validation tests.}
  \label{table-valparam}
\end{table}

\medskip\noindent
{\it 3) Fracture training data set ($\mathbb{MD}^{\text{Frac}}_{{\text{train}}}$, with 1 dynamic MD simulation sample):} The domain of the graphene sheet is set as a square: $[-50 \AA, 50 \AA] \times [-50 \AA, 50 \AA]$.  
The MD grid initially contains a slit (edge crack) of length $25 \AA$ oriented vertically extending from the lower surface. 
The vertical edges of the MD grid have prescribed velocities in the $x_1$ direction that tend to open the crack.
To help maintain stable crack growth, the prescribed velocities decrease linearly with $x_2$, thus tending to limit
the crack growth velocity.
A schematic of the crack pattern at the $40$-th time step in the MD simulation can be find in Figure \ref{fig-mddispl}(c). For the purpose of validation on different grid resolutions, the density, displacement and external loading are computed at two sets of coarse-grained nodes, which are spaced 5\AA\, or 3.17\AA, apart on a square lattice, respectively.

\medskip\noindent
{\it 4) Fracture test data set ($\mathbb{MD}^{\text{Frac}}_{{\text{test}}}$, with 1 dynamic MD simulation sample):} 
To demonstrate that the learned material model applies to different loading scenarios and crack patterns, one additional test case is considered. Here, the MD region and the pre-existing slit are the same as in the fracture training data set. 
However, instead of hard loading along the vertical edges,
a non-zero body force is applied  to the atoms in the MD grid as:
\[
  b_1=b_0\big[ e^{-t/t_r}(1-e^{-t/t_r})\big]
  \sin\left(\frac{\pi x_1}{L}\right)
  e^{-(1/2+x_2/L)}
\]
where $L=100 \AA$ is the edge length of the sample,
$t_r$ is a constant pulse duration time,
and $b_0$ is a positive constant.
(The origin is at the center of the sample.)
This loading exerts a pulse that tends to open the crack.
The resulting crack pattern, which includes branching, is substantially different from that which occurs in the training data.  
A view of the crack pattern at the $40$-th time step in the MD simulation can be found in Figure \ref{fig-mddispl}(d).

As metrics of accuracy on tests, we compare the prediction from the learnt peridynamics model with the ground-truth data from coarse-grained MD measurements. Solution contours are provided as a qualitative validation. With the purpose of providing a quantitative comparison, we also calculate the averaged (in time) mean square errors (MSEs) of the displacement field and the damage field. To provide a fair comparison between different sets, all these qualitative accuracy metrics are normalized with respect to the ground-truth data.

\begin{figure}
\centering
\includegraphics[width=.48\textwidth]{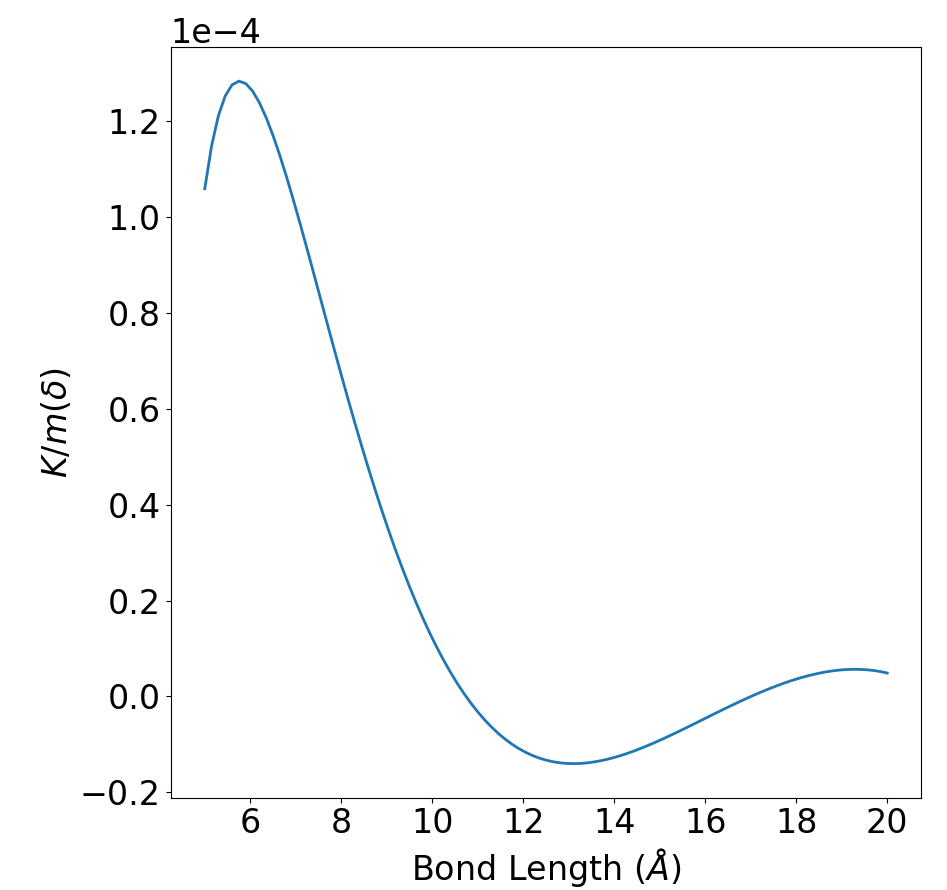}
\includegraphics[width=.48\textwidth]{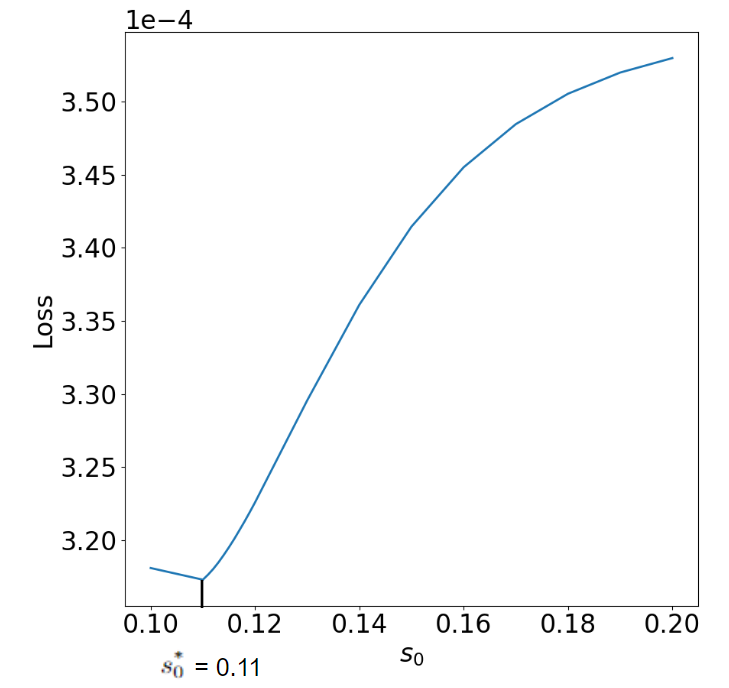}
\caption{Learning results on a single layer graphene sheet. Left: The optimal influence function $K$ for the LPS model. Right: The optimal damage criterion $s_0^*$ is obtained at $0.11$.}
\label{fig:opt_kernel_s0}
\end{figure}

For the kernel learning step, we have followed a similar procedure as in \cite{you2022data}. The learned influence function $K$ is plotted in the left plot of Figure \ref{fig:opt_kernel_s0}, and the optimal material parameter are obtained as 
$\lambda = -0.4796 ({\rm TPA})$, $\mu = 0.7978 ({\rm TPA})$, with the Poisson's ratio $\nu = -0.4297$, and the horizon size $\delta = 20.0 \AA$. 
Then, for the damage criterion learning step, since the crack initiates at the $5$-th time steps, we use the fracture training data set from the $5$-th time step till the $20$-th time step to learn the damage criterion $s_0$, then solve for the optimal $s_0$ by minimizing the loss $\widetilde{\rm Res}(\mcT^{\text{Frac}})$ in \eqref{eqn:optim}. 
Note that when calculating the loss function, we apply Dirichlet-type boundary conditions on a layer of particles near the boundary of our square domain, and hence only the particles in $[-50 \AA + 2\delta, 50 \AA -2\delta] \times [-50 \AA + 2\delta, 50 -2\delta]$ are considered in \eqref{eqn:optim}. This setting differs from the settings in non-fracture data sets, where periodic boundary conditions are considered for all samples. This is due to the fact that it is generally non-realistic to prescribe the periodic boundary condition in the problem with a crack, since the crack itself does not satisfy the periodic condition. This fact also highlights the generalizability of the proposed approach: our homogenized surrogate model can handle data sets with different domains, loadings, and also boundary conditions.
A demonstration of the loss function for different values of $s_0$ is provided in the right plot of Figure \ref{fig:opt_kernel_s0}. The optimal damage criterion is obtained as $s_0^* = 0.11$, which is consistent with the results $s_0 = 0.145$ inferred directly from MD data set in \cite{silling2022peridynamic}.

\subsection{Extrapolation to Longer Time Simulations}\label{sec:test_long}

\begin{figure}
    \centering
    \includegraphics[width=1.\textwidth]{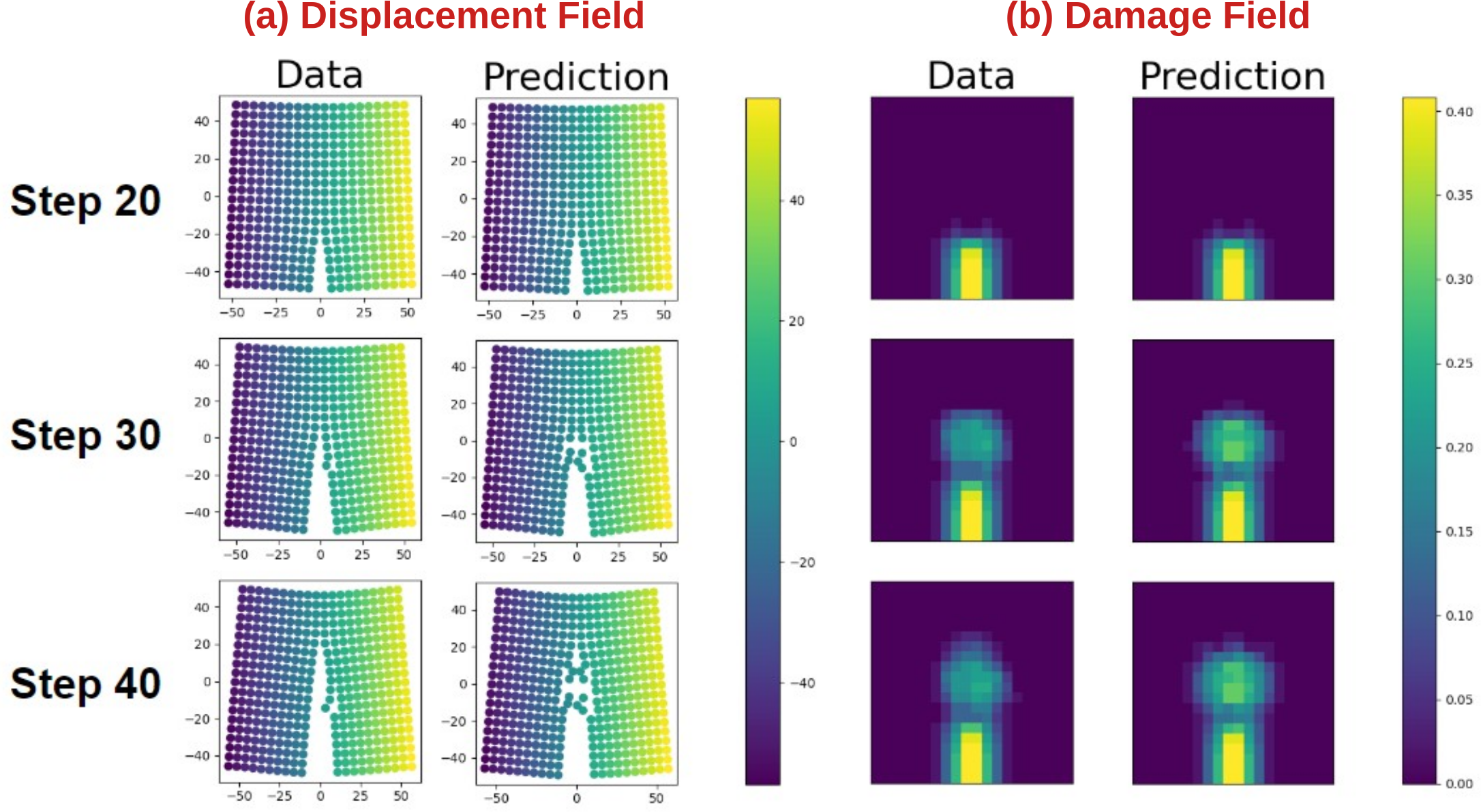}
    \caption{Comparison of the prediction and the ground truth measurement from the MD data set at time steps 20, 30, and 40, on the {\it fracture training data set} where the graphene sheet is subject to zero body force. Here, the first 20 steps were used for training, then we use the learnt model to predict for the next 20 steps. (a) Comparison on the displacement field, where the color of the particles represents the horizontal displacement. (b) Comparison on the damage field.}
    \label{fig:crackzero}
\end{figure}

Next, we validate the learnt model, by using it in a longer term simulation on the fracture training data set, to predict the material deformation and crack propagation upto the $40$-th step. Note that we have used the data upto the $20$-th time step for the purpose of training, and therefore this test can be seen as an investigation on the long-term extrapolation capability of our coarse-grained surrogate model. 
To solve for the displacement field from LPS model, at the $n$-th time step, we first assume there is no broken bond and solve for the displacement field $\hat{\ub}^{n+1}$, then we update the bond-stretch for each connecting bond, and we keep solving for the displacement until there is no new bond breaking. Then, we define the damage profile at each particle $\xb_i$ at time step $n$ as 
\begin{equation}\label{eqn:damage}
    \phi(\xb_i)^n = 1 - \frac{\sum_{\xb_j \in B_{\delta}(\xb_i)} \gamma^n(\xb_i,\xb_j)}{\sum_{\xb_j \in B_{\delta}(\xb_i)} 1}. 
\end{equation}
Figure \ref{fig:crackzero} shows the comparison of displacement and damage fields at time steps 20, 30, and 40. It is observed that the prediction not only matches the data within the training set (step 20) but also exhibits a good agreement at steps 30 and 40, which are not included in the training set. This result suggests that our learned damage criterion $s_0$ is applicable to longer term simulations out of the training data set. For the first 40 steps, we have obtained $27\%$ relative error for the prediction of displacement field and $9\%$ relative error for damage field.




\subsection{Generalization to Different Body Forces and Crack Patterns}\label{sec:test_general}

\begin{figure}
    \centering
    \includegraphics[width=1.\textwidth]{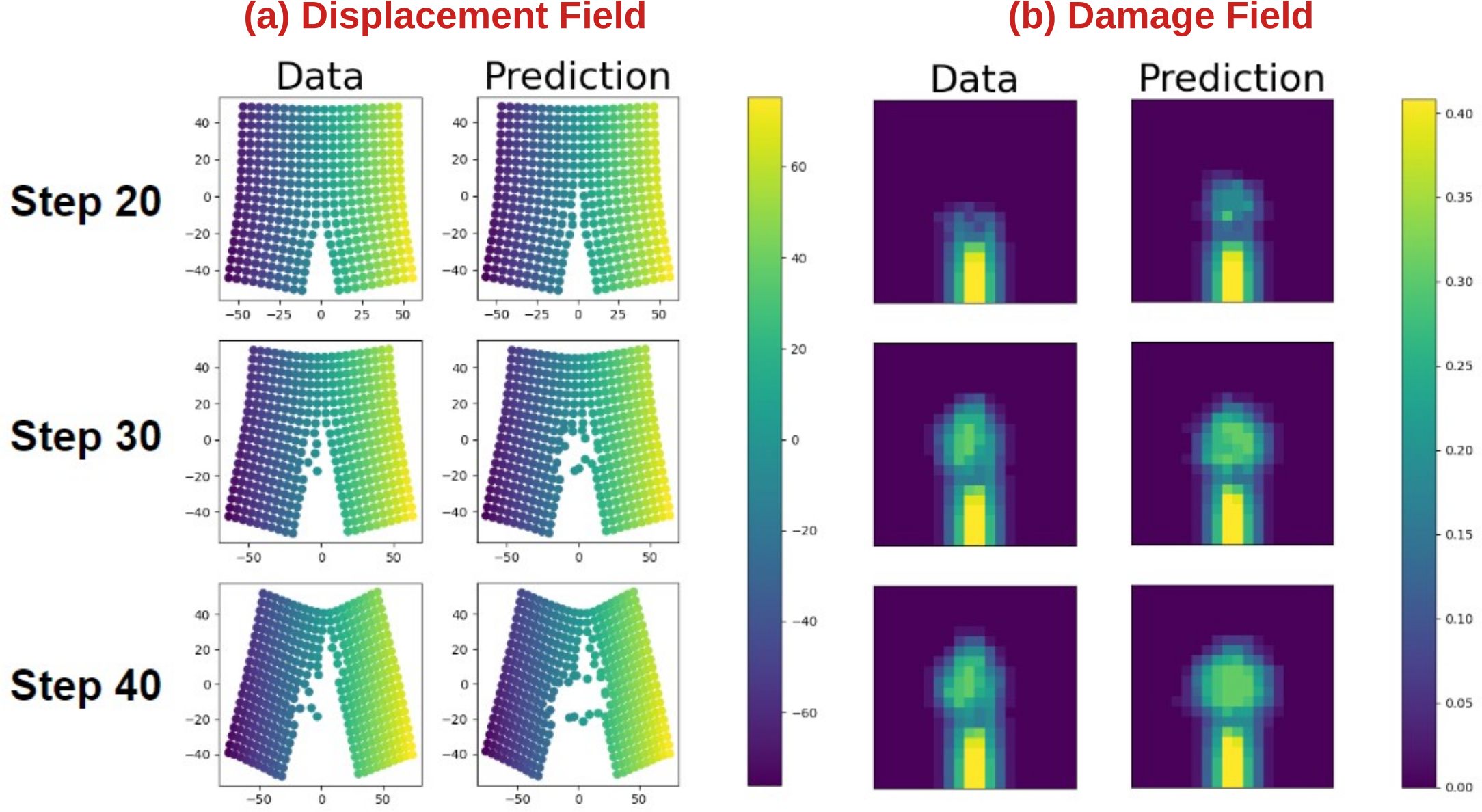}
    \caption{Comparison of the prediction and the ground truth measurement from the MD data set at time steps 20, 30, and 40, on the {\it fracture test data set} where the graphene sheet is subject to unseen and nonzero body force. (a) Comparison on the displacement field, where the color of the particles represents the horizontal displacement. (b) Comparison on the damage field.}
    \label{fig:cracknonzero}
\end{figure}

In this Section, we use the learned LPS surrogate to model the same graphene sheet subject to a different body force load as described in the fracture test data set. Differs from the settings in the training data set, in this data set the graphene sheet is subject to nonzero body load, with its crack pattern at the $40$-th time step illustrated Figure \ref{fig-mddispl}(d). Compared with the crack pattern in the training data set (see Figure \ref{fig-mddispl}(c)), the crack path in this test data set is less symmetric and bifurcates at the middle of the domain. Hence, with this example we not only investigate the extrapolation capability of the learnt model by making a longer time (40 steps) predictions, but also aim to verify its generalizability, since both the loading scenario and crack pattern from this test data set are not covered in the training data. All these factors make the validation more challenging. In Figure \ref{fig:cracknonzero} we show the prediction of displacement and damage fields from the learnt LPS model upto time step $40$. Visually good agreements are observed between the coarse-grained data and LPS predictions. 
This example has qualitatively validated that the learned material damage model can be directly applied to problems with different body force. For the first 40 steps, we have obtained $58\%$ relative error for the prediction of displacement field and $15\%$ relative error for damage field.




\subsection{Generalization to Different Resolutions}\label{sec:test_fine}

\begin{figure}
    \centering
    \includegraphics[width=1.\textwidth]{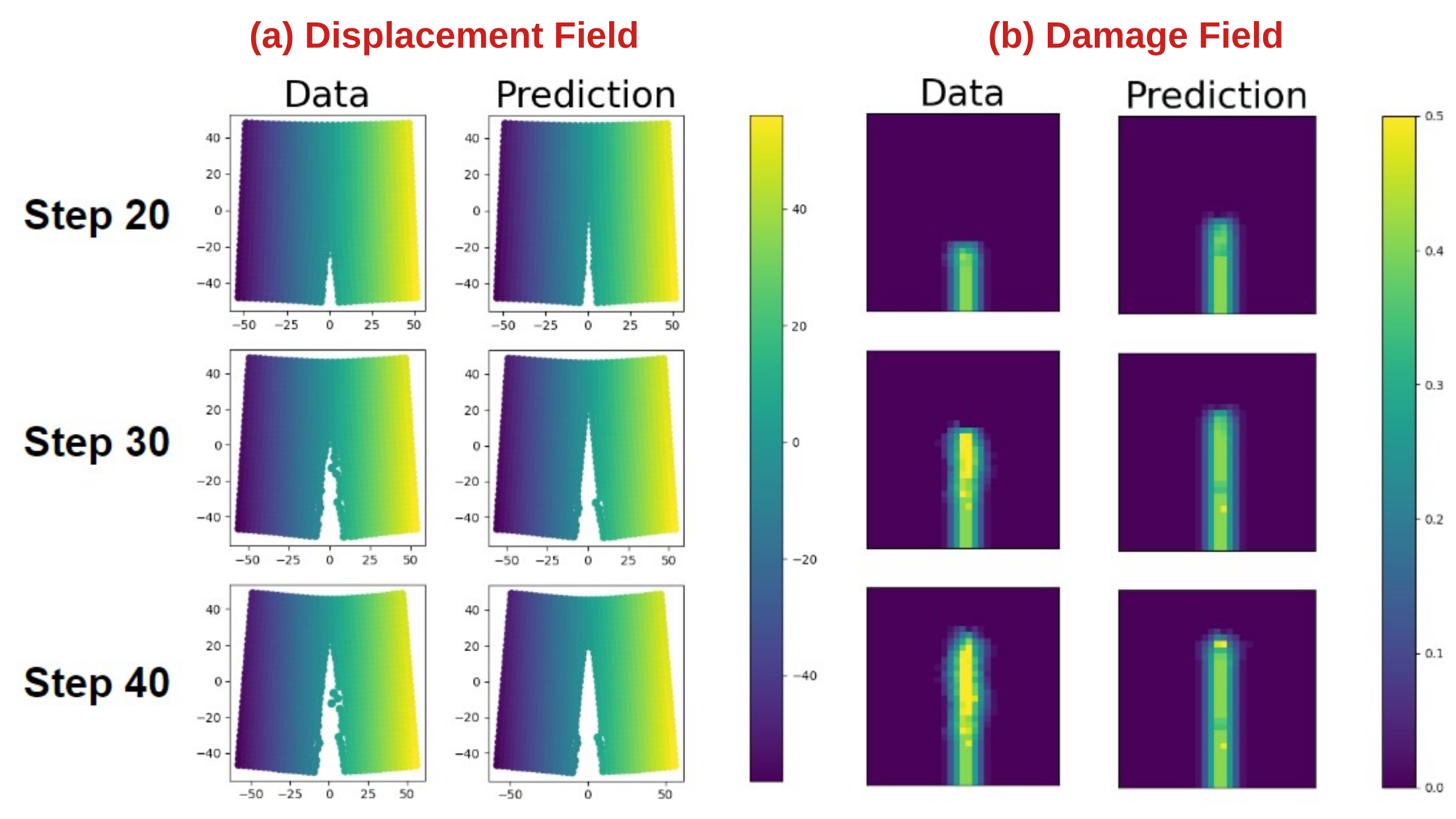}
    \caption{Comparison of the prediction and the ground truth measurement from the MD data set at time steps 20, 30, and 40, on the {\it fracture training data set with fine grids} where the graphene sheet is subject to zero body force. Here, we used coarser grids data in the first 20 steps for training, then we use the learnt model to predict for the next 20 steps on a finer resolution. (a) Comparison on the displacement field, where the color of the particles represents the horizontal displacement. (b) Comparison on the damage field.}
    \label{fig:crackzero_fine}
\end{figure}

Last but not least, we study the resolution generalizability of our learning algorithm. Specifically, we use the same MD data as the training data, but evaluate the density, displacement and force loading on a coarse-grained grid with smaller grid size $h=3.17 \AA$. Since all training data sets are with a fixed grid size $h=5 \AA$, with this study we aim to investigate if the learnt surrogate model allows the grid size to be rescaled, providing a multiscale capability and allowing for flexible solver resolution and reductions in computational cost. As suggested by \cite{you2022data}, we scale the horizon size $\delta$ proportionally with the grid size $h$ to provide a fixed horizon/grid size ratio. In particular, we take $\delta = 4h = 12.68 \AA$. Then, the optimal damage criterion is also scaled correspondingly to guarantee a consistent critical release rate. 
As proved in \cite{zhang2018state}, the damage criterion and horizon size should satisfy the relation $s_0 \propto \frac{1}{\sqrt{\delta}}$ in the LPS model. Thus we use $s_0 := 0.11\sqrt{\frac{5}{3.17}}$ for our fine scale simulation. In Figure \ref{fig:crackzero_fine} we show the displacement and and damage field prediction results upto time step 40, demonstrating a qualitative agreement between the coarse-grained MD data and our numerical predictions. On the displacement field, we have obtained $19\%$ relative error in average for the first 40 steps, which is even smaller than the prediction error without resolution alternation on the same data set ($27\%$ as shown in Section \ref{sec:test_long}). For the damage field, one can see that the crack pattern predicted by our surrogate model grows faster than the crack from MD data set. Therefore, a larger prediction error, $30\%$ average for damage field in the first 40 steps, is obtained. 
This example suggests that the surrogate model can provide qualitatively consistent displacement predictions on different resolutions. On the other hand, the prediction on damage field is sub-optimal, possibly due to the fact that the material crack originates from microscale phenomena, and hence is more sensitive to the prediction scales. To improve the prediction accuracy on the damage field across different resolutions, practitioners might consider performing the damage criterion learning step on the new resolution, to provide a correction for the damage criterion.



\section{Conclusions}\label{sec:conclusion}

In this paper, we demonstrate a data-driven workflow to extract a coarse-grained surrogate model from MD data with fracture. Firstly, to handle the discontinuities induced by material fracture in the MD displacement measurements, a smoothness indicator function is introduced, to automatically choose the locally smoothest stencil from the neighborhood of each coarse grained grid. As such, the coarse-graining measurements are built based on this adaptive stencil, to automatically handle the discontinuities in MD displacement data set without overly smoothing the crack pattern. It is shown that this novel adaptive procedure significantly improves the capability of capturing the location of crack interfaces. Then, based on the coarse-grained data set we proposed to extract a peridynamics surrogate, which is a continuum mechanics model that allows a natural treatment of discontinuities by replacing spatial derivatives of stress tensors with integrals of force density functions. By learning the kernel function of the integral and the damage criterion with a two-step optimization approach, we obtain a linear peridynamic solid model which provides good agreement with nanoscale test data while being capable to provide further material deformation and fracture predictions under unseen domain settings, loading scenarios, and even different grid resolutions. These features greatly reducing the cost of the calculation in comparison with MD, especially when used together with different discretization resolutions.

Although the present work focuses on relatively small deformations and a linear peridynamics model, the results suggest that this method may impact a broader range of materials and applications. As another natural follow-up work, one may further combine the nonlocal model with the approximation power of neural networks, to obtain a nonlinear peridynamics model in the form of integral neural operators \cite{kovachki2021neural,you2022physics,you2022nonlocal,you2022learning}.

\section*{Acknowledgements}

H. You and Y. Yu would like to acknowledge support by the National Science Foundation under award DMS-1753031 and
the AFOSR grant FA9550-22-1-0197. Portions of this research were conducted on Lehigh University's Research Computing infrastructure partially supported by NSF Award 2019035. 
 
S. Silling and M. D'Elia would like to acknowledge the support of the Sandia National Laboratories (SNL) Laboratory-directed Research and Development program and by the U.S. Department of Energy (DOE), Office of Advanced Scientific Computing Research (ASCR) under the Collaboratory on Mathematics and Physics-Informed Learning Machines for Multiscale and Multiphysics Problems (PhILMs) project. This article has been authored by an employee of National Technology and Engineering Solutions of Sandia, LLC under Contract No. DE-NA0003525 with the U.S. Department of Energy (DOE). The employee owns all right, title and interest in and to the article and is solely responsible for its contents. The United States Government retains and the publisher, by accepting the article for publication, acknowledges that the United States Government retains a non-exclusive, paid-up, irrevocable, world-wide license to publish or reproduce the published form of this article or allow others to do so, for United States Government purposes. The DOE will provide public access to these results of federally sponsored research in accordance with the DOE Public Access Plan https://www.energy.gov/downloads/doe-public-access-plan.


\bibliographystyle{elsarticle-num}
\bibliography{snl}

\end{document}